\documentclass[amsmath,amssymb,twocolumn,nofootinbib]{revtex4-1}
\usepackage{graphicx}
\usepackage{epsfig}
\usepackage{color}

\begin{document}
\title{Renormalization group  inspired autonomous equations \\for secular effects in de Sitter space}
\author{Alexander~Yu.~Kamenshchik}
\email{kamenshchik@bo.infn.it}
\affiliation{Dipartimento di Fisica e Astronomia, Universit\`a di Bologna\\ and INFN,  Via Irnerio 46, 40126 Bologna,
Italy,\\
L.D. Landau Institute for Theoretical Physics of the Russian
Academy of Sciences,\\
 Kosygin street. 2, 119334 Moscow, Russia}
\author{Tereza Vardanyan}
\email{tereza.vardanyan@aquila.infn.it}
\affiliation{Dipartimento di Fisica e Chimica, Universit\`a di L'Aquila, 67100 Coppito, L'Aquila, Italy\\ and 
INFN, Laboratori Nazionali del Gran Sasso, 67010 Assergi, L'Aquila, Italy}

\begin{abstract}
We develop a method for treating a series of secularly growing terms obtained from quantum perturbative calculations: autonomous first-order differential equations are constructed such that they reproduce this series to the given order. The exact solutions of these equations are free of secular terms and approach a finite limit at late times. This technique is illustrated for the well-known problem of secular growth of correlation functions of a massless scalar field with a quartic self-interaction in de Sitter space. For the expectation value of the product of two fields at coinciding space-time points we obtain a finite late-time result that is very close to the one following from Starobinsky's stochastic approach.\\\\\\\
\end{abstract}
%\pacs{98.80.Jk, 98.80.Cq, 04.20.-q, 04.20.Jb}
\maketitle

\section{Introduction}
The renormalization group (RG), born in the framework of quantum field theory, has become one of its most efficient tools (see, e.g., the reviews \cite{Shirkov1, Shirkov} and references therein). The origin of this concept is connected with the fact that removal of ultraviolet divergences leads to some arbitrariness in defining the renormalized parameters of the theory. However, physics should not be affected by this arbitrariness: observable quantities must be independent of the renormalization scale. Using this requirement, combined with the information obtained from perturbation theory, we can derive differential equations whose solutions are equivalent to partial resummation of the perturbative series.        

Later it became clear that the area of application of RG ideas is much wider than the problem of renormalization and ultraviolet divergences in quantum field theory. One can mention, first of all, Wilson's version of the renormalization group, which  played an important role in the physics of condensed matter and found further applications in quantum field theory as well \cite{Wilson}.

More recently, some new methods allowing application of RG approach to classical problems of mathematical physics were developed (see e.g. review \cite{Shirkov2} and references therein). For example, in paper \cite{dynamical} it was shown how one can improve the naive perturbative solutions of some rather complicated differential equations. Namely, the authors developed the so-called dynamical renormalization group method by considering differential equations that involve a small parameter and whose zeroth order solutions are bounded functions, while the first iteration reveals a presence of secularly growing terms. These terms spoil the validity of the perturbative expansion past a certain point in time; in order to deal with them, an arbitrary intermediate time scale is introduced and the initial conditions are renormalized. The RG equations for the renormalized initial conditions can be derived from the fact that the intermediate time scale does not appear in the original problem. The solutions of these RG equations allow to improve the original perturbative result by extending its domain of validity.

Sometimes one encounters situations where the ``traditional'' ultraviolet and infrared divergences are intertwined with secular effects. This can happen when we consider a quantum field theory set in an expanding background. An interesting example is the de Sitter spacetime represented in the system of coordinates with flat spatial sections (Poincar\'e patch). Here the infrared divergences are much stronger than in Minkowski spacetime and different kinds of secular effects arise \cite{Emil, Prokopec:2007ak}. For a massless minimally coupled scalar field there is a secular growth already present at the level of the free theory: the long-wavelength part of the expectation value $\langle \phi^2(\vec x,t) \rangle$ evaluated in the Bunch-Davies vacuum \cite{vacuum,vacuum1,vacuum2} grows linearly with time \cite{grow,grow1,grow2,grow3,grow4}. If there is a self-interaction of the type $\lambda \phi^4$, the perturbative calculation of the long-wavelength part of $\langle \phi^2(\vec x,t) \rangle$ gives a series with terms that behave like $\lambda^n (Ht)^{2n+1}$. When $Ht>1/\sqrt\lambda$, the perturbation theory breaks down, so it can't make reliable predictions at late times. 

A remarkable non-perturbative technique for calculating the late-time expectations values was proposed by Starobinsky in \cite{Star} and further developed in many papers, in particular in \cite{Star-Yoko}. In paper \cite{Star-Yoko} it was suggested that the dynamics of the long-wavelength modes of the quantum field $\phi(\vec x, t)$ can be described by a classical stochastic variable whose probability distribution satisfies a Fokker-Planck type equation. The authors showed that at late times any solution of this equation approaches the static solution, which in turn can be used to calculate the expectation values. In essence, Starobinsky's Fokker-Planck equation manages to resum the leading secular terms of the perturbative expansion \cite{Woodard}. The emergence of the stochastic picture from the full quantum evolution of the theory was presented in more recent works \cite{Tereza,Burgess:2014eoa,Burgess:2015ajz}. 

Knowing how efficient the RG methods are, it is tempting to try to apply them to the secular effects in de Sitter space.  An interesting attempt, based on the dynamical renormalization group method, was undertaken in the thought-provoking paper \cite{Burgess}. However, the obtained results do not reproduce those known from the stochastic approach.

In the present paper we develop a semi-heuristic method for taking the late-time limit of a series of secularly growing terms obtained from quantum perturbative calculations. Namely, we construct autonomous first-order differential equations such that our perturbative results can be obtained from these equations by simple iterations. In the series we consider, even the zeroth order term grows secularly with time, but when we construct an autonomous equation that reproduces this series to linear order in the coupling constant, its exact solution approaches a finite limit at late times. Applied to $\langle \phi^2(\vec x,t) \rangle$ of $\phi^4$-theory in de Sitter spacetime, this procedure gives a result that coincides with the Hartree-Fock approximation. 

To see if we can improve this result, we build an autonomous equation that reproduces the perturbative series to second order. This equation is more complicated: it can be integrated, but in general it is not possible to write its exact solution as an explicit function of time. As we shall explain in the next section, in some cases we can look for the explicit solution in the form of perturbative expansion in a parameter that characterizes the deviation of this solution from the solution of the previous, simpler autonomous equation. At late times the function obtained in this way approaches a finite limit, and in the case of $\langle \phi^2(\vec x,t) \rangle$ this finite value is very close to that known from the stochastic approach.     

The structure of the paper is the following: in the second section we present our method in rather general terms; in the third section we use it to calculate the asymptotic values for $\langle \phi^2(\vec x,t) \rangle$ and $\langle \phi^4(\vec x,t) \rangle$ of $\phi^4$-theory in de Sitter space and compare the results with the stochastic approach; the last section contains concluding remarks. In Appendices we present perturbative calculations of the leading secular terms in the two- and four-point functions.
  
\section{Autonomous equations inspired by renormalization group}
Let us consider the following problem. We are looking for a function $f(t)$, which is an expectation value of an operator; it depends on time and a small parameter $\lambda$. 
We do not have the dynamical equation governing this function, but we have some information obtained by perturbative methods. We know that when the parameter $\lambda$ is very small, the function has the following form 
\begin{equation}
f(t) = A(t-t_0)-\lambda B (t-t_0)^3+\mathcal O(\lambda^2)\;,
\label{pert}
\end{equation}
where $A$ and $B$ are some positive constants. As $t-t_0$ grows, the perturbation theory breaks down and the expansion \eqref{pert} can no longer be trusted. Even when $\lambda = 0$, the function $f = A(t-t_0)$ grows linearly with time, and it is difficult to use the dynamical renormalization group method \cite{dynamical}, which works quite well when the zeroth order approximation is a bounded function. At the same time  we know (or we can guess from some physical considerations) that as $t\to\infty$, the function $f$ should approach a constant value. How can we model this behavior and follow what happens at late times? 

Our suggestion is the following: we shall try  to find a simple autonomous first-order differential equation that produces the first two terms of the expression (\ref{pert}) by iterations.
Namely, at zeroth order we have 
\begin{equation}
f(t) = A(t-t_0)\;,
\label{zero}
\end{equation}
and this function can be obtained as a solution of a simple differential equation 
\begin{equation}
\frac{df}{dt} = A\;.
\label{zero1}
\end{equation}
Now we would like to generate the second term on the right-hand side of Eq. (\ref{pert}) by iteration of an autonomous first-order differential equation. To do this, it is enough to add to the right-hand side of the differential equation (\ref{zero1}) the term $-\lambda\frac{3B}{A^2}f^2$, so that we have the following equation:
\begin{equation}
\frac{df}{dt} = A -\lambda\frac{3B}{A^2}f^2\;.
\label{equation}
\end{equation}
Solving this equation by iterations up to first order in $\lambda$ we find the expression (\ref{pert}). 

We can also obtain Eq.~\eqref{equation} in a slightly different way: notice that \eqref{pert} can be represented as
\begin{eqnarray}
f(t)=y(t)-\lambda{B\over A^3}\left[y(t)\right]^3+\mathcal O(\lambda^2)\;,
\label{pert1}
\end{eqnarray}
where $y(t)$ is the zeroth order term,
\begin{eqnarray*}
y(t)\equiv A(t-t_0)\;.
\end{eqnarray*}
Differentiating \eqref{pert1} with respect to $t$, we get  
\begin{eqnarray*}
{df\over dt}=A-\lambda{3B\over A^2}y^2+\mathcal O(\lambda^2)\;.
\end{eqnarray*}
Within the given accuracy, $y^2$ on the right side of this equation can be replaced by $f^2$; hence, we arrive at 
\begin{eqnarray}
\frac{df}{dt} = A -\lambda\frac{3B}{A^2}f^2,
\label{equation1}
\end{eqnarray}
which coincides with Eq.~\eqref{equation}. Fortunately, this equation is integrable and its solution is 
\begin{equation}
f(t) = \sqrt{\frac{A^3}{3\lambda B}}\tanh\left[\sqrt{\frac{3\lambda B}{A}}(t-t_0)\right]\;,
\label{solution}
\end{equation}
where the integration constant is chosen such that $f(t_0)=0$. It is easy to see that expanding \eqref{solution} in powers of $\lambda$, we reproduce the first two terms of Eq.~\eqref{pert}. 
The remarkable feature of this expression is that it is regular for all values of $t$, and when $t\to\infty$, one has 
\begin{equation*}
f(t) \to \sqrt{\frac{A^3}{3\lambda B}}\;.
\end{equation*}
Another interesting feature of this solution is its non-analyticity with respect to the small parameter $\lambda$. Note that appearance of non-analyticity can also be observed rather often when the dynamical renormalization group is used \cite{dynamical}. 
 
In principle, this procedure can be generalized for the situation when we have more than two terms coming from perturbation theory. Suppose that we know our function $f$ up to the quadratic in $\lambda$ term:
\begin{equation}
f(t) = A(t-t_0)-\lambda B (t-t_0)^3 +\lambda^2 C (t-t_0)^5+\mathcal O(\lambda^3)\;.
\label{more}
\end{equation}
Rewriting everything in terms of the zeroth order term,
\begin{eqnarray*}
f(t)=y(t)-\lambda{B\over A^3}\left[y(t)\right]^3+\lambda^2{C\over A^5}\left[y(t)\right]^5+\mathcal O(\lambda^3)\;,
\end{eqnarray*}
and taking the time derivative, we obtain
\begin{eqnarray}
\frac{df}{dt} = A - \lambda\frac{3B}{A^2}y^2 +\lambda^2\frac{5C}{A^4}y^4+\mathcal O(\lambda^3)\;.
\label{equation2}
\end{eqnarray}
To the given order, $y^4$ on the right side can be replaced by $f^4$. To replace $y^2$ we notice that   
\begin{eqnarray*}
f^2=y^2-2\lambda{B\over A^3}y^4+\mathcal O(\lambda^2)\;,
\end{eqnarray*}
so, to $\lambda^2$-order, Eq.~\eqref{equation2} can be written as
\begin{equation}
\frac{df}{dt} = A - \lambda\frac{3B}{A^2}f^2 +\lambda^2\left(\frac{5C}{A^4}-\frac{6B^2}{A^5}\right)f^4\;.
\label{more1}          
\end{equation}
This equation is also integrable and we can obtain its implicit solution
\begin{equation}
t = t(f)\;.
\label{time}
\end{equation}
The exact form of \eqref{time} depends on the sign of the determinant of the right side of Eq.~\eqref{more1}, and in general it is not possible to find the explicit form of $f(t)$. But in some cases we can obtain the solution of Eq.~\eqref{more1} in the form of perturbative expansion in a small parameter. 

What would this small parameter be? Looking at Eq.~\eqref{more1}, we see that if the coefficients in the expansion \eqref{more} are such that\begin{eqnarray} 
C={6B^2\over5A}\;,
\label{HF}
\end{eqnarray}
then the coefficient of the quartic term $f^4$ is equal to zero, and we are back to Eq.~\eqref{equation1} and its solution \eqref{solution}. This is not surprising, since the expansion of \eqref{solution} up to $\lambda^2$-order gives \eqref{more}, with the coefficient $C$ that satisfies the condition \eqref{HF}:         
\begin{equation*}
f(t) = A(t-t_0)-\lambda B (t-t_0)^3 +\lambda^2 {6B^2\over 5A} (t-t_0)^5\;.
\end{equation*}
Let us now split the actual $C$ in the following way:
\begin{eqnarray*}
C={6B^2\over 5A}+\Delta C={6B^2\over 5A}(1+\epsilon)\;,
\end{eqnarray*}
where
\begin{eqnarray}
\epsilon\equiv{5A\over6B^2}\Delta C={5AC\over6B^2}-1\;.
\label{epsilon}
\end{eqnarray}
With the above notations, Eq.~\eqref{more1} can be rewritten as
\begin{eqnarray*}
\frac{df}{dt} = A - \lambda\frac{3B}{A^2}f^2 +\lambda^2\epsilon\frac{6B^2}{A^5}f^4\;.
\end{eqnarray*}
If we rescale $f(t)$, 
\begin{eqnarray*}
F(t)\equiv\sqrt{3\lambda B\over A^3}f(t)\;,
\end{eqnarray*}
our differential equation will have the following form
\begin{eqnarray}
\frac{dF}{dt} = \sqrt{3\lambda B\over A}\left(1 - F^2 + {2\over3}\epsilon F^4\right)\;.
\label{Fequation}
\end{eqnarray}
We see that if $\epsilon$ is small, we can look for the solution of this equation in the form of the perturbative expansion  
\begin{eqnarray}
F(t)=F_0(t)+\epsilon F_1(t)+\mathcal{O}(\epsilon^2)\;.
\label{F1epsilon}
\end{eqnarray}
The zeroth order term satisfies the equation
\begin{eqnarray*}
\frac{dF_0}{dt} =\sqrt{3\lambda B\over A}\left(1 - F_0^2\right)\;,
\end{eqnarray*}
and its solution, with $F_0(t_0)=0$, is
\begin{eqnarray*}
F_0(t)=\tanh\left[\sqrt{\frac{3\lambda B}{A}}(t-t_0)\right]\;.
\end{eqnarray*}
For the first-order term we have
\begin{eqnarray*}
\frac{dF_1}{dt} = \sqrt{3\lambda B\over A}\left(-2F_0F_1 +{2\over3}F_0^4\right)\;,
\end{eqnarray*}
and its solution, with $F_1(t_0)=0$, is
\begin{widetext}
\begin{eqnarray*}
F_1(t)={1\over3}\tanh\left[\sqrt{3\lambda B\over A}(t-t_0)\right]+{{2\over3}\tanh\left[\sqrt{3\lambda B\over A}(t-t_0)\right]-\sqrt{3\lambda B\over A}(t-t_0)\over\cosh^2\left[\sqrt{3\lambda B\over A}(t-t_0)\right]}\;.
\end{eqnarray*}
Hence, to first-order in $\epsilon$, our original function $f(t)$ is given by
\begin{eqnarray}
\!\!f(t)=\left(1+{\epsilon\over3}\right)\sqrt{\frac{A^3}{3\lambda B}}\tanh\left[\sqrt{\frac{3\lambda B}{A}}(t-t_0)\right]+{\epsilon\left\{{2\over3}\sqrt{\frac{A^3}{3\lambda B}}\tanh\left[\sqrt{3\lambda B\over A}(t-t_0)\right]-A(t-t_0)\right\}\over\cosh^2\left[\sqrt{3\lambda B\over A}(t-t_0)\right]}\;,
\end{eqnarray}
\end{widetext}
and as $t\to\infty$, it approaches the limit  
\begin{eqnarray}
f(t)\to\sqrt{\frac{A^3}{3\lambda B}}\left(1+{\epsilon\over3}\right)=\sqrt{\frac{A^3}{3\lambda B}}\left(\frac23+\frac{5}{18}\frac{AC}{B^2}\right)\;,~~~~
\label{asymp2}
\end{eqnarray}
where we used the definition of $\epsilon$ \eqref{epsilon} to obtain the last equality.

Let us also consider a function, whose perturbative expansion has a slightly different secular behavior 
\begin{equation}
g(t) = J(t-t_0)^2-\lambda K(t-t_0)^4+\lambda^2 L(t-t_0)^6+\mathcal O(\lambda^3)\;.
\label{new}
\end{equation}
If we only keep the first two terms, the corresponding autonomous first-order differential equation that reproduces them is 
\begin{equation}
\frac{dg}{dt} = 2\sqrt{Jg} - \lambda\frac{3K}{J^{3/2}}g^{3/2}\;.
\label{final}
\end{equation}
The solution of this equation is
\begin{eqnarray}
g(t)={2J^2\over3\lambda K}\tanh^2\left[\sqrt{\frac{3\lambda K}{2J}}(t-t_0)\right]\;,
\label{tanh2}
\end{eqnarray}
and at large times it tends to the finite limit 
\begin{equation}
g(t)\to {2J^2\over3\lambda K}\;.
\end{equation} 
With a little more work we can find the autonomous equation that reproduces all three terms in \eqref{new},  
\begin{equation}
\frac{dg}{dt} = 2\sqrt {Jg} - \lambda\frac{3K}{J^{3/2}}g^{3/2} +\lambda^2\left({5L\over J^{5/2} }-{17\over4}\frac{K^2}{J^{7/2}}\right)g^{5/2}\;.
\label{final}
\end{equation}
To find an approximate solution of this equation, we proceed similarly to the previous case (cf. Eq~\eqref{more1}). In the expansion of \eqref{tanh2} in powers of $\lambda$ the coefficient proportional to $\lambda^2$ is equal to $17K^2/20 J$; if we parameterize the relative difference between this coefficient and the coefficient $L$,           
\begin{eqnarray}
\tilde{\epsilon}\equiv{20J\over17K^2}\left(L-{17K^2\over20 J}\right)={20JL\over17K^2}-1\;
\end{eqnarray}
and also rescale the function $g(t)$ 
\begin{eqnarray}
G(t)\equiv{3\lambda K\over2J^2}g(t)\;,
\end{eqnarray}
we obtain the following equation
\begin{eqnarray}
\frac{dG}{dt} = \sqrt{6\lambda B\over J}\left(G^{1/2} - G^{3/2} + {17\over18}\tilde{\epsilon} G^{5/2}\right)\;.
\end{eqnarray}
If $\tilde{\epsilon}$ is small, we can solve this differential equation perturbatively. Solving it to first order in $\tilde{\epsilon}$ and going back to the function $g(t)$, we find that at late times it approaches the limit 
\begin{equation}
g(t)\to {2J^2\over3\lambda K}\left(1+{17\over18}\tilde{\epsilon}\right)={2J^2\over3\lambda K}\left({1\over18}+{20\over18}{JL\over K^2}\right)\;.
\label{glimit}
\end{equation}

\section{Resummation of secular terms in de Sitter space} 

We shall consider the de Sitter spacetime represented as an expanding spatially flat Friedmann universe with the following metric 
\begin{equation}
ds^2=dt^2-a^2(t)\delta_{ij}dx^idx^j\;,
\label{dS}
\end{equation}
where the scale factor $a(t)$ is 
\begin{equation}
a(t) = e^{Ht}\;.
\label{a}
\end{equation}
Here $t$ is a cosmic time coordinate and $H$ is the Hubble constant or the inverse of the de Sitter radius. The cosmic time in an expanding de Sitter universe runs in the interval $-\infty < t < \infty$. It will also be convenient to use a conformal time coordinate $\eta$, which is related to the cosmic time $t$ by the condition $dt=a(\eta)d\eta$. Expressed in terms of the conformal time, the metric is 
\begin{equation}
ds^2 = a^2(\eta)(d\eta^2-\delta_{ij}dx^idx^j)\;,
\label{conformal}
\end{equation}
where 
\begin{equation}
a(\eta) = -\frac{1}{H\eta}\;,
\label{conformal1}
\end{equation}
and $\eta$ runs from $-\infty$ to $0$. 

We shall consider a massless minimally coupled scalar field with a quartic self-interaction. Its  action is 
\begin{equation}
S = \int d^4 \sqrt{-g}\left(\frac12 g^{\mu\nu}\partial_{\mu}\phi\partial_{\nu}\phi-\frac{\lambda}{4}\phi^4\right)\;.
\label{action}
\end{equation}
The Klein-Gordon equation for the free (non-interacting) field is 
\begin{equation}
\ddot{\phi}(\vec{x},t)+3H\dot{\phi}(\vec{x},t)-\frac{\nabla^2}{a^2}\phi(\vec{x},t) = 0\;,
\label{KG}
\end{equation}
where ``dot'' stands for the derivative with respect to the cosmic time and $\nabla^2$ is the three-dimensional Laplacian.
Making the Fourier transformation and the transition to the conformal time, we can rewrite Eq. (\ref{KG}) as follows:
\begin{equation}
\phi_k''(\eta)-\frac{2}{\eta}\phi_k'(\eta)+k^2
\phi_k(\eta) = 0\;,
\label{KG1}
\end{equation}
where $k = |\vec k|$ and ``prime'' denotes the derivative with respect to the conformal time.
The solutions of this equation have the form
\begin{equation}
\phi_k(\eta) \sim (1\pm ik\eta) e^{\mp ik\eta}\;.
\label{BD}
\end{equation}
Now, $\phi$ can be decomposed as 
\begin{equation}
\phi(\vec{x},t) = \int {d^3\vec{k}\over (2\pi)^3}\left\{u_k(\eta)e^{i\vec{k}\cdot\vec{x}}a_{\vec{k}}+u_k^*(\eta)e^{-i\vec{k}\cdot\vec{x}}a_{\vec k}^\dagger\right\}\;,
\label{BD1}
\end{equation}
 where $a$ and $a^\dagger$ are the annihilation and creation operators while $u$ and $u^*$ are basis functions proportional to the solutions 
 (\ref{BD}).
The choice of the function $u$ defines the choice of the creation and annihilation operators (which, naturally, should satisfy the standard commutation relations),
which, in turn, defines the vacuum state. If one wants to have a vacuum that in the remote past $\eta\to-\infty$ (or, equivalently, for modes with very short physical wavelength, $-kH\eta\gg H$) behaves like the vacuum in Minkowski spacetime, one should choose 
\begin{equation}
u_k(\eta) = {iH\over\sqrt{2k^3}}(1+ik\eta)e^{-ik\eta}\;.
\label{BD2}
\end{equation}
Such a choice is called the Bunch-Davies vacuum \cite{vacuum,vacuum1,vacuum2}. 

For small values of the physical momentum, $-k\eta\ll1$, the two-point correlator constructed from the mode functions \eqref{BD2} behaves like $1/k^3$, in contrast to flat spacetime, where it behaves like $1/k^2$, which means that the infrared divergences are stronger in de Sitter spacetime. Let us take a closer look at this correlator: at the level of the free theory the equal-time two-point function is given by
\begin{eqnarray}
\!\!\!\!\!\!\langle\phi(\vec x,t)\phi(\vec y,t)\rangle_{\lambda^0}&=&\int\frac{d^3{\vec k}}{(2\pi)^3}u_k(\eta)u_k^*(\eta)e^{i\vec k\cdot(\vec x-\vec y)}\nonumber \\
&=&{H^2\over2}\int\frac{d^3{\vec k}}{(2\pi)^3}{(1+k^2\eta^2)\over k^3}{e^{i\vec k\cdot(\vec x-\vec y)}}\;.\nonumber\\
\label{free1}
\end{eqnarray}
We would like to find the late-time behavior of the {\it long-wavelength} part of \eqref{free1}, that is, the part coming from the modes with physical momenta much less than the Hubble scale, $-k\eta\ll1$. In the case of coinciding spatial points, we obtain (the subscript $L$ stands for ``long-wavelength part'')
\begin{eqnarray}
\!\!\!\langle\phi^2(\vec x,t)\rangle_{\lambda^0,L}&=&{H^2\over4\pi^2}\int_{\kappa}^{-1/\eta}{dk\over k}(1+k^2\eta^2)\nonumber\\
&=&-{H^2\over4\pi^2}\bigg(\ln{(-\kappa\eta)}-{1\over2}+{\kappa^2\eta^2\over2}\bigg)\;,
\label{late}
\end{eqnarray}
where we introduced an infrared cutoff $\kappa$ for the comoving momentum $k$, since the integral is divergent at $k=0$. For $t\to\infty$ (i.e., $-\kappa\eta\ll1$), the first term in \eqref{late} dominates, so in the late-time limit we have 
\begin{eqnarray}
\langle\phi^2(\vec x,t)\rangle_{\lambda^0,L}={H^3\over4\pi^2}(t-t_0)\;,
\label{late0}
\end{eqnarray}  
where $t_0\equiv(1/H)\ln(\kappa/H)$; thus, it grows linearly with time \cite{grow,grow1,grow2,grow3,grow4}.

In the presence of the quartic self-interaction, the expression (\ref{late0}) will receive perturbative corrections. In Appendix I we used the ``in-in'' (Schwinger - Keldysh) formalism \cite{Schwinger,Keldysh,Bakshi:1962dv,Jordan,Weinberg:2005vy} to evaluate these corrections. Because this formalism involves four types of propagators, the calculations are rather cumbersome; however, it is still possible to extract the leading late-time behavior of $\langle \phi^2 (\vec{x},t) \rangle_L$: to second order in $\lambda$, it is given by the sum of \eqref{late0}, \eqref{2point1} and \eqref{2point2},
\begin{equation}
\langle \phi^2 (\vec{x},t) \rangle_L = \frac{H^3}{4\pi^2}(t-t_0) - \lambda\frac{H^5}{24\pi^4}(t-t_0)^3+\lambda^2\frac{H^7}{80\pi^6}(t-t_0)^5.
\label{Onemli}
\end{equation}
Results similar to \eqref{Onemli} were also presented in the series of works \cite{On-Wood,Brunier,Kahiya,Onemli}. 

We can identify the expression (\ref{Onemli}) with the general expression for the function $f(t)$, introduced in the preceding section (cf. Eq. (\ref{more})). Then the coefficients $A, B$ and $C$ for $f(t) =\langle \phi^2 (\vec{x},t) \rangle_L$  are 
\begin{equation}
A = \frac{H^3}{4\pi^2}\;,
\label{A}
\end{equation}
\begin{equation}
B = \frac{H^5}{24\pi^4}\;,
\label{B}
\end{equation}
\begin{equation}
C =\frac{H^7}{80\pi^6}\;.
\label{C}
\end{equation}

Let us first consider the autonomous equation arising when we take into consideration only the first two terms in \eqref{Onemli}, 
\begin{equation}
\frac{d\langle \phi^2 (\vec{x},t) \rangle_L}{dt} = \frac{H^3}{4\pi^2} -  \frac{2\lambda}{H}\langle \phi^2 (\vec{x},t) \rangle_L^2\;.
\label{equation10}
\end{equation}
The exact solution of this equation, with the initial condition $\langle \phi^2 (\vec{x},t_0) \rangle_L=0$, is 
\begin{equation}
\langle \phi^2 (\vec{x},t) \rangle_L = \frac{H^2}{\sqrt{8\lambda}\pi}\tanh\left[\sqrt{\frac{H^2\lambda}{2\pi^2}}(t-t_0)\right]\;.
\label{solution10}
\end{equation}
The secular growth disappears and at late times we have 
\begin{equation}
\langle \phi^2 (\vec{x},t) \rangle_L \to \frac{H^2}{\sqrt{8\lambda}\pi}\;.
\label{solution11}
\end{equation}

If all three terms in expression (\ref{Onemli}) are taken into account, then the corresponding autonomous equation is
\begin{eqnarray}
\!\!\!\frac{d\langle \phi^2 (\vec{x},t) \rangle_L}{dt}\!=\! \frac{H^3}{4\pi^2} -  \frac{2\lambda}{H}\langle \phi^2 (\vec{x},t) \rangle_L^2+\frac{16\pi^2\lambda^2}{3H^5}\langle \phi^2 (\vec{x},t) \rangle_L^4\;.\nonumber\\
\label{equation20}
\end{eqnarray}
Following what we did in the previous section, we can solve it perturbatively in the parameter $\epsilon$ defined in \eqref{epsilon}. To first order in $\epsilon$, the solution asymptotically approaches \eqref{asymp2}, so with the values of $A,B$ and $C$ from (\ref{A})--(\ref{C}), we obtain\begin{equation}
\langle \phi^2 (\vec{x},t) \rangle_L\to\frac{H^2}{\sqrt{8\lambda}\pi}\left(1+{\epsilon\over3}\right)=\frac{7}{6}\frac{H^2}{\sqrt{8\lambda}\pi}\;.
\label{solution21}
\end{equation}
This asymptotic value is $7/6$ times greater than the asymptotic value \eqref{solution11} obtained from the simpler autonomous equation \eqref{equation10}. 
 
We shall compare our results for the asymptotic behavior of $\langle \phi^2 (\vec{x},t) \rangle_L$ to the ones obtained in the Hartree-Fock approximation as well as in the stochastic approach \cite{Star,Star-Yoko}. Following paper \cite{Star-Yoko}, let us write the equation of motion for our scalar field with the action \eqref{action},
\begin{eqnarray}
\phi_{;\mu}^{;\mu}=-\lambda\phi^3\;.
\end{eqnarray}
Multiplying both sides by $\phi$, integrating the left side by parts and taking expectation values of the field operators results in     
\begin{equation}
\frac12\langle \phi^2\rangle_{;\mu}^{;\mu}-\langle \phi^{,\mu}\phi_{,\mu}\rangle =- \lambda \langle \phi^4 \rangle\;.
\label{equation4}
\end{equation}
Using the Hartree-Fock (Gaussian) approximation, $\langle \phi^4 \rangle = 3\langle \phi^2 \rangle^2$, for the term on the right-hand side, Eq.~\eqref{equation4} can be written as
\begin{equation}
\frac12\langle \phi^2\rangle_{;\mu}^{;\mu}-\langle \phi^{,\mu}\phi_{,\mu}\rangle =- 3\lambda \langle \phi^2 \rangle^2\;.
\end{equation}
When $\lambda=0$, the dominant contribution to the infrared part ($-k\eta\ll1$) of the left-hand side comes from the $3H\partial/\partial t$ part of the d'Alembertian; from \eqref{late0} we see that
\begin{eqnarray}
{d\langle \phi^2\rangle_L\over\ dt}={H^3\over4\pi^2}\;,
\end{eqnarray}
so it can be concluded that
\begin{equation}
\frac{d\langle \phi^2 \rangle_L}{dt} =  \frac{H^3}{4\pi^2} -  \frac{2\lambda}{H}\langle \phi^2 \rangle_L^2\;.
\label{equation5}
\end{equation}
This equation coincides with our equation \eqref{equation10}, and, naturally, their exact solutions and asymptotic behavior also coincide. This means that by using the perturbative expansion of $\langle \phi^2 (\vec{x},t) \rangle_L$ to first order in $\lambda$ and constructing the autonomous equation, we reproduce the results obtained in the Hartree-Fock approximation. 

The stochastic approach argues \cite{Star-Yoko} that the behavior of the long-wavelength part of the quantum field $\phi(\vec x,t)$ can be modeled by an auxiliary  classical stochastic variable $\varphi$ with a probability distribution $\rho(\varphi,t)$ that satisfies the Fokker-Planck equation   
\begin{equation}
{\partial\rho\over\partial t} 
= {H^3\over8\pi^2}{\partial^2\rho\over\partial\varphi^2} 
+ {1\over3H}{\partial\over\partial\varphi}
\biggl({\partial V\over\partial\varphi} \rho(t,\varphi) \biggr)\;,
\label{F-P}
\end{equation}
where $V(\varphi)=\lambda\varphi^4/4$; namely the expectation value of any quantity constructed from the long-wavelength part of $\phi(\vec x,t)$ is equal to the expectation value of the same quantity constructed from  the variable $\varphi$.    

At late times any solution of the equation \eqref{F-P} approaches the static solution
\begin{equation}
\rho(\varphi) = \left(\frac{32\pi^2\lambda}{3}\right)^{\frac14}\frac{1}{\Gamma\left(\frac14\right)H}\exp\left(-\frac{2\pi^2\lambda\varphi^4}{3H^4}\right)\;.
\label{equation22}
\end{equation}
Using this distribution, one can calculate the expectation value of $\varphi^2$:
\begin{eqnarray}
\langle \varphi^2 \rangle = \int_{-\infty}^{\infty} d\varphi\, \varphi^2 \rho(\varphi)=\sqrt{\frac{3}{2\pi^2}}\frac{\Gamma\left(\frac34\right)}{\Gamma\left(\frac14\right)}\frac{H^2}{\sqrt{\lambda}}\;.
\label{equation23}
\end{eqnarray}
Now we can compare this with our result \eqref{solution21} obtained by using the autonomous equation \eqref{equation20}:
\begin{equation}
\frac{\langle \varphi^2 \rangle-\langle \phi^2 \rangle_L}{\langle \varphi^2 \rangle}\approx0.0036=0.36\%\;.
\label{comparison}
\end{equation}
We see that our method gives a result that is extremely close to the value obtained in the stochastic approach. 

There is one caveat that we would like to mention. In order to get the asymptotic value \eqref{asymp2} (and, hence, \eqref{solution21}), we had to assume that the parameter $\epsilon$ is small, so that the expansion \eqref{F1epsilon} is a good approximation to the exact solution of Eq.~\eqref{Fequation}. If we use the values of $A,B$ and $C$ from (\ref{A})--(\ref{C}), we obtain that for the $\phi^4$ theory in de Sitter space this parameter is equal to $1/2$, which can hardly be considered very small as compared to $1$. Nonetheless, as we just saw, the asymptotic value of $\langle \phi^2 (\vec{x},t) \rangle_L$ produced by this approximation is surprisingly close to the one from the stochastic picture. At this point we can note that such a situation is not uncommon when one works with perturbation theory. Indeed, rather often we treat various parameters in a way as if they were very small and still obtain some reasonable results.           

At the end of this section we also consider the application of our method to the calculation of $\langle \phi^4 \rangle_L$. One can calculate the perturbative expression for this correlator: to second order in $\lambda$ its leading late-time behavior is given by the sum of \eqref{4point0}, \eqref{4point1} and \eqref{four2} (see Appendices),
\begin{eqnarray}
\langle \phi^4(\vec x,t) \rangle_L = {3H^6\over16\pi^4}(t-t_0)^2&&-\lambda {3H^8\over 32\pi^6}(t-t_0)^4\nonumber\\&&+\lambda^2{53 H^{10}\over960\pi^8}(t-t_0)^6\;.
\label{four1}
\end{eqnarray}
The structure of the expression (\ref{four1}) coincides with that presented in Eq.~(\ref{new}) at the end of the preceding section; the coefficients $J$, $K$ and $L$ are
\begin{equation}
J = \frac{3H^6}{16\pi^4}\;,
\label{G}
\end{equation} 
\begin{equation}
K = \frac{3H^8}{32\pi^6}\;,
\label{K}
\end{equation}
\begin{equation}
L = \frac{53H^{10}}{960\pi^8}\;.
\label{L}
\end{equation}
Using \eqref{glimit} with the appropriate values of coefficients, we conclude that in the limit $t \to \infty$,
\begin{equation}
\langle \phi^4(\vec x,t) \rangle_L\to\frac{H^4}{4\pi^2\lambda}\left(1+{17\over18}\tilde{\epsilon}\right)=\frac{221}{648\pi^2}{H^4\over\lambda}\;.
\label{four3}
\end{equation}
On the other hand, using the stationary probability distribution (\ref{equation22}) found from the Fokker-Planck equation, we can calculate this expectation value as 
\begin{eqnarray}
\langle \varphi^4 \rangle = \int_{-\infty}^{\infty} d\varphi\,\varphi^4 \rho(\varphi)= \frac{3H^4}{8\pi^2\lambda}.
\label{four4} 
\end{eqnarray}
Comparing \eqref{four3} with its stochastic counterpart,
\begin{equation}
\frac{\langle \varphi^4\rangle-\langle \phi^4\rangle_L}{\langle \varphi^4 \rangle}\approx0.0905=9.05\%\;,
\end{equation}
we see that the difference is bigger than in the case of $\langle \phi^2 \rangle_L$. 

To explain why the result we obtained for $\langle \phi^2 \rangle_L$ is so much closer to its stochastic value than the result for $\langle \phi^4\rangle_L$, let us look back at expression \eqref{solution21} for the late-time value of $\langle \phi^2 \rangle_L$. The factor in front of the parentheses is the asymptotic value we obtained by solving the lower-order autonomous equation, that is, the autonomous equation that reproduces the perturbative series to first order in $\lambda$; this value is already fairly close to the stochastic value \eqref{equation23}: the difference is about $15\%$. By contrast, in the case of $\langle \phi^4 \rangle_L$ the asymptotic value obtained from the lower-order autonomous equation, which is equal to the factor in front of the parentheses in \eqref{four3}, is farther away from the stochastic value \eqref{four4}: the difference is about $33\%$. This means that in the case of $\langle \phi^4\rangle_L$ more iterations are needed to get closer to the stochastic value: an autonomous equation reproducing the perturbative series to third (or higher) order in $\lambda$ should be considered.

\section{Concluding remarks}
Many quantum field theories set in an expanding background have secularly growing terms in their perturbatively calculated correlation functions. In the case of the massless minimally coupled scalar field in de Sitter space this growth manifests itself even at zeroth order, thereby making it difficult to apply the renormalization group methods. In this paper we presented a technique for taking the late-time limit of this type of perturbative series by constructing auxiliary autonomous first-order differential equations. By applying this technique to $\phi^4$-theory in de Sitter spacetime, we calculated the late-time limit of expectation values of products of two and four fields at coinciding space-time points and compared our results with those obtained from the stochastic approach. In principle, the method developed here can also be used to calculate the late-time limit of correlators of fields at different space-time points. 

It would be interesting to explore the subdominant secular terms present in perturbative series, that is, the terms that are suppressed by additional powers of the coupling constant with respect to the leading secular terms. If we retain these subdominant terms in the perturbative expansion, the autonomous equations needed to reproduce this expansion will, of course, change. How will the late-time limit of their solutions differ from the ones obtained with only leading secular terms? We hope to address this question in future work.    

\section*{Acknowledgements}

We are grateful to A. A. Starobinsky for fruitful discussions. The work of A.K. was partially supported by the RFBR grant No 18-52-45016. 

\section*{Appendix I: Perturbative calculations of correlators using the ``in-in'' formalism}
\subsection{The ``in-in'' formalism}
To calculate first- and second-order corrections to the two-point function, as well as the four-point function, we will work in the interaction picture and use the ``in-in'' formalism \cite{Schwinger,Keldysh,Bakshi:1962dv,Jordan,Weinberg:2005vy}. In this formalism equal-time n-point functions can be written as 
\begin{widetext}
\begin{eqnarray}
&&\langle\phi(t,\vec x_1)\cdots \phi(t,\vec x_n)\rangle=\bigl\langle 0\big| 
U_I^\dagger(t,-\infty)\phi_I(t,\vec x_1)\cdots \phi_I(t,\vec x_n) 
U_I(t,-\infty)\big|0\bigr\rangle
\nonumber\\
&&
~~~=\bigl\langle 0\big| 
\bigl(\bar{T} e^{i\int_{-\infty}^t dt'\, H_I(t')}\bigr) 
\phi_I(t,\vec x_1)\cdots \phi_I(t,\vec x_n) 
\bigl(T e^{-i\int_{-\infty}^t dt'\, H_I(t')}\bigr) \big|0\bigr\rangle,
\label{in}
\end{eqnarray}
\end{widetext}
where $\phi_I(t,\vec x_n)$, just as in ``in-out'' formalism, are interaction picture fields with time evolution governed by the free theory Hamiltonian; $H_I$ is the interaction Hamiltonian in the interaction picture; $T$ stands for time-ordering, $\bar{T}$ --- for anti-time-ordering; $|0\rangle$ is the vacuum state of the free theory, which, as explained in section 3 of the main text, is chosen to be the Bunch-Davies state. In what follows we suppress the subscript $I$ of the interaction picture fields. 

There are some differences between this formalism and the usual ``in-out'' formalism, which is used for calculations of scattering processes. For scattering precesses we start from some initial ``in''-state at $t=-\infty$, evolve it from $t=-\infty$ to $t=\infty$, and then calculate its overlap with a final ``out''-state at $t=\infty$: both the initial and final states of the system are specified. In the case of ``in-in'' formalism \eqref{in} only the initial state is specified: both the ``in''-state $|0\rangle$ and its Hermitian conjugate $\langle 0|$ are evolved from $-\infty$ to a time $t$ with $U_I(t,-\infty)$ and $U_I^\dagger(t,-\infty)$  respectively, then the product of fields is sandwiched between them. 

There is also another way to interpret \eqref{in}: we start with the initial state at $-\infty$, evolve forward to a time $t$, where the product of fields is inserted, then evolve backwards to $-\infty$. This is why the ``in-in'' formalism is also called ``closed-time-path'' formalism. This interpretation makes it possible to write \eqref{in} in terms of a single time-ordered expression \cite{Jordan,Collins:2011mz}: label the fields on the forward-flowing part of the path with a ``$+$'' superscript, the fields on the backward-flowing part of the path --- with a ``$-$'' superscript; thereby, \eqref{in} can be written as                           
\begin{widetext}
\begin{eqnarray}
\langle\phi(t,\vec x_1)\cdots \phi(t,\vec x_n)\rangle=\bigl\langle0\big| 
T\bigl(\phi^+(t,\vec x_1)\cdots \phi^+(t,\vec x_n)e^{-i\int_{-\infty}^t dt'\;[H_I^+(t')-H_I^-(t')]}\bigr)\big|0\bigr\rangle,
\label{sc} 
\end{eqnarray}
\end{widetext}
where $H_I^{\pm}(t)\equiv H_I[\phi^{\pm}(t,\vec{x})]$, and the time-ordering operation is extended in the following way: two ``$+$'' fields are ordered as usual,
\begin{eqnarray*}
T(\phi^+(t,\vec{x})\phi^+(t',\vec{y}))&=&\Theta(t-t')\phi^+(t,\vec{x})\phi^+(t',\vec{y})\nonumber\\&&+\Theta(t'-t)\phi^+(t',\vec{y})\phi^+(t,\vec{x})\;,
\end{eqnarray*}
``$-$'' fields always occur after ``$+$'' fields,
\begin{eqnarray*}
T(\phi^+(t,\vec{x})\phi^-(t',\vec{y}))&=&\phi^-(t',\vec{y})\phi^+(t,\vec{x})\;,\\
T(\phi^-(t,\vec{x})\phi^+(t',\vec{y}))&=&\phi^-(t,\vec{x})\phi^+(t',\vec{y})\;,
\end{eqnarray*}
and two ``$-$'' fields are ordered in the opposite of the usual sense,
\begin{eqnarray*}
T(\phi^-(t,\vec{x})\phi^-(t',\vec{y}))&=&\Theta(t'-t)\phi^-(t,\vec{x})\phi^-(t',\vec{y})\nonumber\\&&+\Theta(t-t')\phi^-(t',\vec{y})\phi^-(t,\vec{x})\;.\nonumber
\end{eqnarray*}
We can use Wick's theorem to express the time-ordered products in \eqref{sc} in terms of field contractions, but unlike in ``in-out'' formalism, there are four types of Wick contractions (and hence, four propagators)
\begin{eqnarray*}
&&\langle 0|T(\phi^+(t,\vec{x})\phi^+(t',\vec{y}))|0\rangle\\
&&~~=\Theta(t-t')G^{>}(t,\vec x;t',\vec y)+\Theta(t'-t)G^{<}(t,\vec x;t',\vec y)\;,\\
&&\langle 0|T(\phi^+(t,\vec{x})\phi^-(t',\vec{y}))|0\rangle=G^{<}(t,\vec x;t',\vec y)\;,\\
&&\langle 0|T(\phi^-(t,\vec{x})\phi^+(t',\vec{y}))|0\rangle=G^{>}(t,\vec x;t',\vec y)\;,\\
&&\langle 0|T(\phi^-(t,\vec{x})\phi^-(t',\vec{y}))|0\rangle\\
&&~~=\Theta(t'-t)G^{>}(t,\vec x;t',\vec y)+\Theta(t-t')G^{<}(t,\vec x;t',\vec y)\;,
\end{eqnarray*}
where $G^{>}(t,\vec x;t',\vec y)$ and $G^{<}(t,\vec x;t',\vec y)$ are Wightman functions\\
\begin{eqnarray*}
G^{>}(t,\vec x;t',\vec y)&=&\langle 0|\phi(t,\vec{x})\phi(t',\vec{y})|0\rangle\\
&=&\int\frac{d^3\vec{k}}{(2\pi)^3}\;e^{i{\vec{k}(\vec{x}-\vec{y})}}G_k^{>}(t,t')\;,\\
G^{<}(t,\vec x;t',\vec y)&=&\langle 0|\phi(t',\vec{y})\phi(t,\vec{x})|0\rangle\\
&=&\int\frac{d^3\vec{k}}{(2\pi)^3}\;e^{i{\vec{k}(\vec{x}-\vec{y})}}G_k^{<}(t,t')\;.
\end{eqnarray*}
The Wightman functions associated with the Bunch-Davies vacuum are
\begin{eqnarray}
G_k^>(t,t')&=&u_k(\eta)u_k^*(\eta')\nonumber \\
&=&\frac{H^2}{2k^3}(1+ik\eta)(1-ik\eta')e^{-ik(\eta-\eta')}\nonumber\\
G_k^<(t,t')&=&u_k^*(\eta)u_k(\eta')\nonumber \\
&=&\frac{H^2}{2k^3}(1-ik\eta)(1+ik\eta')e^{ik(\eta-\eta')}\;. \label{W}
\end{eqnarray}
\subsection{The two-point function and mass counterterm at first order in $\lambda$}
For the scalar field with quartic self-interaction, the interaction Hamiltonian is 
\begin{eqnarray*}
H_I(t)=\int d^3\vec{x}\;\left(a^3(t){\lambda\over4}\phi^4-\mathcal L_c\right)\;,
\label{HI}\;
\end{eqnarray*}
where $\mathcal L_c$ is the counterterm Lagrangian required to renormalize the theory \eqref{action},  
\begin{eqnarray*}
\mathcal L_c=a^3(t)\left({\delta_1\over2}\dot\phi^2-{\delta_2\over 2a^2(t)}(\partial_i\phi)^2-{\delta_m\over2}\phi^2-{\delta_{\lambda}\over4}\phi^4\right)\;.
\end{eqnarray*}
From \eqref{sc} the first-order correction to $\langle\phi(\vec x,t)\phi(\vec y,t)\rangle$ is given by
\begin{eqnarray*}  
&&\langle\phi(\vec x,t)\phi(\vec y,t)\rangle_{\lambda}\nonumber\\
&&~~=-i\int_{-\infty}^t dt'\;\langle0|T(\phi^+(\vec x,t)\phi^+(\vec y,t)[H_I^+(t')-H_I^-(t')])|0\rangle.
\end{eqnarray*}
Diagrammatically, it is the sum of one-loop and counterterm diagrams shown on Fig.1 ($\delta_1$, $\delta_2$ and $\delta_{\lambda}$ are equal to zero at first order in $\lambda$). 
\setlength{\unitlength}{1mm}
\thicklines
\begin{picture}(90,30)
\put(0,15){\circle*{2}}
\put(-4,11){$(\vec{x},t)$}
\put(1,15){\line(1,0){20}}
\put(22,15){\circle*{2}}
\put(22,21.5){\circle{13}}
\put(21,11){$t'$}
\put(23,15){\line(1,0){20}}
\put(44,15){\circle*{2}}
\put(52,15){\circle*{2}}
\put(40,11){$(\vec{y},t)$}
\put(52,15){\line(1,0){20}}
\put(72,14){$\bigotimes$}
\put(75,15){\line(1,0){20}}
\put(95,15){\circle*{2}}
\put(52,11){}
\put(90,11){}
\put(75,11){}
\put(72,18){$\delta_m$}
\put(48,5){Fig. 1}
\end{picture}
Contractions that correspond to the one-loop diagram give
\begin{widetext}
\begin{eqnarray}
12\bigg({-i\lambda\over 4}\bigg)\int\frac{d^3{\vec k}}{(2\pi)^3}{e^{i\vec k\cdot(\vec x-\vec y)}}\int_{-\infty}^{\eta}{d\eta'}a^4(\eta')\bigg[(G^>_k(\eta,\eta'))^2-(G^<_k(\eta,\eta'))^2\bigg]\int\frac{d^3{\vec \ell}}{(2\pi)^3}G^>_\ell(\eta',\eta'),
\label{1loop}
\end{eqnarray}
and for the counterterm diagram we have
\begin{eqnarray}
2\bigg({-i{\delta_m}\over 2}\bigg)\int\frac{d^3{\vec k}}{(2\pi)^3}{e^{i\vec k\cdot(\vec x-\vec y)}}\int_{-\infty}^{\eta}{d\eta'}a^4(\eta')\bigg[(G^>_k(\eta,\eta'))^2-(G^<_k(\eta,\eta'))^2\bigg],
\end{eqnarray}
\end{widetext}
where we switched from the cosmic time $t$ to the conformal time $\eta$, and $a(\eta')=1/(-H\eta')$. The loop integral in \eqref{1loop}
\begin{eqnarray}
&&\int\frac{d^3{\vec \ell}}{(2\pi)^3}G^>_\ell(\eta',\eta')
={H^2\over 2}\int\frac{d^3{\vec \ell}}{(2\pi)^3}{(1+\ell^2\eta'^2)\over\ell^3}\label{1loop'}
\end{eqnarray}
has both infrared (IR) and ultraviolet (UV) divergencies. To regulate them we introduce IR and UV cutoffs, with the UV cutoff set at a fixed {\it physical} momentum $\Lambda_{UV}$ and the IR cutoff set at a fixed {\it comoving} momentum $\kappa$ (for explanation of this choice see \cite{Xue:2011hm,Senatore:2009cf}): 
\begin{eqnarray}
&&{H^2\over4\pi^2}\int_{\kappa}^{a(\eta')\Lambda_{UV}}{d\ell\over\ell}(1+\ell^2\eta'^2)\nonumber \\
&&={H^2\over4\pi^2}\bigg[\ln{\bigg({\Lambda_{UV}\over\kappa}\bigg)}-\ln{(-H\eta')}+{\Lambda_{UV}^2\over2H^2}\bigg].
\label{1loopreg}
\end{eqnarray}
We can see that in order to absorb the UV-divergencies, the mass counterterm should be equal to   
\begin{eqnarray}
{\delta_m}=-{3\lambda\over4\pi^2}\bigg[{\Lambda_{UV}^2\over 2}+H^2\ln{\bigg({\Lambda_{UV}\over\mu_{UV}}\bigg)}\bigg].
\label{mc}
\end{eqnarray}
Taking this counterterm into account and choosing the UV renormalization scale $\mu_{UV}$ to be equal to $H$, we obtain the UV-renormalized result
\begin{eqnarray}
&&\langle\phi(\vec x,t)\phi(\vec y,t)\rangle_{\lambda}={3i\lambda H^2\over 4\pi^2}\!\!\int\frac{d^3{\vec k}}{(2\pi)^3}{e^{i\vec k\cdot(\vec x-\vec y)}}\nonumber\\&&~\times\int_{-\infty}^{\eta}\!\!{d\eta'}a^4(\eta')\bigg[(G^>_k(\eta,\eta'))^2-(G^<_k(\eta,\eta'))^2\bigg]\ln{(-\kappa\eta')}.\nonumber\\
\label{1looplate}
\end{eqnarray}
The arguments of the exponents in the Wightman functions are products of the momentum and conformal time, ($\pm k(\eta-\eta')$), so in order to perform the time integration we need to change the integration variable from $\eta'$ to $z'\equiv k\eta'$. Because of this, we also need to rewrite the $\ln{(-\kappa\eta')}$ in terms of $z'$ 
\begin{eqnarray}
\ln{(-\kappa\eta')}d\eta'=\bigg(\ln{(-z')}-\ln{{k\over\kappa}}\bigg){dz'\over k}.
\label{split1}
\end{eqnarray}
When performing the time integral in \eqref{1looplate}, the time contour is deformed to have a small imaginary part in order to project out the vacuum state of the interacting theory from the vacuum state of the free theory \cite{Maldacena:2002vr,Peskin:1995ev}. Hence, there are no contributions coming from the lower limit of the integral, where $\eta'=-\infty(1\pm i\epsilon)$ and the oscillatory exponents in the Wightman functions turn into damping exponents. For late times, the contribution of the upper limit of the integral can be obtained by using the following expansion ($\eta\to0$ and $\eta'\to0$), 
\begin{eqnarray}
&&a^4(\eta')\bigg[(G^>_k(0,\eta'))^2-(G^<_k(0,\eta'))^2\bigg]\nonumber \\
&&={i\over 3k^3\eta'}\bigg(1+{2\over 5}k^2\eta'^2+\mathcal{O}(k^4\eta'^4)\bigg)\nonumber\\
&&={i\over 3k^2z'}\bigg(1+{2\over 5}z'^2+\mathcal{O}(z'^4)\bigg).
\label{expand1}
\end{eqnarray}
Using \eqref{split1} and \eqref{expand1} we obtain that, for $\eta\to0$, \eqref{1looplate} goes as
\begin{widetext}
\begin{eqnarray}
\langle\phi(\vec x,t)\phi(\vec y,t)\rangle_{\lambda}\to{\lambda H^2\over 4\pi^2}\!\!\int\frac{d^3{\vec k}}{(2\pi)^3}{e^{i\vec k\cdot(\vec x-\vec y)}\over k^3}\bigg(\ln{{k\over\kappa}}\ln{(-k\eta)}-{1\over 2}\ln^2{(-k\eta)}\bigg),
\end{eqnarray}
and its long-wavelength part, for $\vec x=\vec y$, is
\begin{eqnarray}
\langle\phi^2(\vec x,t)\rangle_{\lambda, L}&=&{\lambda H^2\over 8\pi^4}\!\!\int_{\kappa}^{-1/\eta}\frac{d{k}}{k}\bigg(\ln{{k\over\kappa}}\ln{(-k\eta)}-{1\over 2}\ln^2{(-k\eta)}\bigg)\nonumber\\&=&{\lambda H^2\over24\pi^4}\ln^3{(-\kappa\eta)}=-{\lambda H^5\over24\pi^4}(t-t_0)^3,
\label{2point1}
\end{eqnarray}
where $t_0\equiv(1/H)\ln(\kappa/H)$ as before (cf. Eq.~\eqref{late0}).
\subsection{$\lambda^2$-correction}
Expanding \eqref{sc} to second order and taking all possible field contractions, we obtain several diagrams with different topologies that contribute to $\langle\phi(\vec x,t)\phi(\vec y,t)\rangle$ at $\lambda^2$-order.
\subsubsection{Two independent loops}
Fig.2 shows the diagram with two independent loops and the appropriate counterterm diagrams that should be combined with it.

\begin{picture}(250,30)
\put(0,15){\circle*{2}}
\put(-4,11){$(\vec{x},t)$}
\put(1,15){\line(1,0){10}}
\put(12,15){\circle*{2}}
\put(12,20){\circle{10}}
\put(12,11){$t'$}
\put(12,15){\line(1,0){12}}
\put(24,15){\circle*{2}}
\put(24,20){\circle{10}}
\put(24,11){$t''$}
\put(24,15){\line(1,0){10}}
\put(34,15){\circle*{2}}
\put(30,11){$(\vec{y},t)$}
\put(40,15){\circle*{2}}
\put(40,11){}
\put(40,15){\line(1,0){10}}
\put(50,15){\circle*{2}}
\put(50,20){\circle{10}}
\put(50,11){}
\put(50,15){\line(1,0){10}}
\put(60,14){$\bigotimes$}
\put(60,18){$\delta_m$}
\put(60,11){}
\put(63,15){\line(1,0){10}}
\put(73,15){\circle*{2}}
\put(73,11){}
\put(79,15){\circle*{2}}
\put(79,15){\line(1,0){10}}
\put(89,14){$\bigotimes$}
\put(89,18){$\delta_m$}
\put(92,15){\line(1,0){10}}
\put(102,15){\circle*{2}}
\put(102,20){\circle{10}}
\put(102,15){\line(1,0){10}}
\put(112,15){\circle*{2}}
\put(79,11){}
\put(89,11){}
\put(102,11){}
\put(112,11){}
\put(118,15){\circle*{2}}
\put(118,11){}
\put(118,15){\line(1,0){10}}
\put(128,14){$\bigotimes$}
\put(128,18){$\delta_m$}
\put(128,11){}
\put(131,15){\line(1,0){10}}
\put(141,14){$\bigotimes$}
\put(141,18){$\delta_m$}
\put(143,15){\line(1,0){10}}
\put(154,15){\circle*{2}}
\put(89,5){Fig. 2}
\end{picture} 
Taking field contractions that correspond to the diagram with two independent loops and making use of the theta functions in the propagators, for this diagram we obtain     
\begin{eqnarray}
&&2(288){1\over 2}\bigg({-i\lambda\over 4}\bigg)^2\int\frac{d^3{\vec k}}{(2\pi)^3}{e^{i\vec k\cdot(\vec x-\vec y)}}\int_{-\infty}^{\eta}{d\eta'}a^4(\eta')\bigg[G^>_k(\eta,\eta')-G^<_k(\eta,\eta')\bigg]\int\frac{d^3{\vec p}}{(2\pi)^3}G^>_p(\eta',\eta')\nonumber \\
&&\times\int_{-\infty}^{\eta'}{d\eta''}a^4(\eta'')\bigg[G^>_k(\eta',\eta'')G^>_k(\eta,\eta'')-G^<_k(\eta',\eta'')G^<_k(\eta,\eta'')\bigg]\int\frac{d^3{\vec \ell}}{(2\pi)^3}G^>_\ell(\eta'',\eta'').
\label{2-loops}
\end{eqnarray}
The diagrams with one loop and one cross (the mass counterterm insertion) give 
\begin{eqnarray}
&&48\bigg({-i\lambda\over 4}\bigg)\bigg({-i{\delta_m}\over 2}\bigg)\int\frac{d^3{\vec k}}{(2\pi)^3}{e^{i\vec k\cdot(\vec x-\vec y)}}\int_{-\infty}^{\eta}{d\eta'}a^4(\eta')\bigg[G^>_k(\eta,\eta')-G^<_k(\eta,\eta')\bigg]\nonumber \\
&&\times\int_{-\infty}^{\eta'}{d\eta''}a^4(\eta'')\bigg[G^>_k(\eta',\eta'')G^>_k(\eta,\eta'')-G^<_k(\eta',\eta'')G^<_k(\eta,\eta'')\bigg]\int\frac{d^3{\vec p}}{(2\pi)^3}\bigg\{G^>_p(\eta',\eta')+G^>_p(\eta'',\eta'')\bigg\},
\label{loop-cr}
\end{eqnarray}
and the diagram with two mass counterterm insertions gives
\begin{eqnarray}
&&2(8){1\over 2}\bigg({-i{\delta_m}\over 2}\bigg)^2\int\frac{d^3{\vec k}}{(2\pi)^3}{e^{i\vec k\cdot(\vec x-\vec y)}}\int_{-\infty}^{\eta}{d\eta'}a^4(\eta')\bigg[G^>_k(\eta,\eta')-G^<_k(\eta,\eta')\bigg]\nonumber \\
&&\times\int_{-\infty}^{\eta'}{d\eta''}a^4(\eta'')\bigg[G^>_k(\eta',\eta'')G^>_k(\eta,\eta'')-G^<_k(\eta',\eta'')G^<_k(\eta,\eta'')\bigg].
\label{2-cross}
\end{eqnarray} 
The sum of \eqref{2-loops} and of the first part of \eqref{loop-cr} (the part with only the first term kept in the curly brackets) gives
\begin{eqnarray}
&&{18\lambda^2H^2\over 4\pi^2}\int\frac{d^3{\vec k}}{(2\pi)^3}{e^{i\vec k\cdot(\vec x-\vec y)}}\int_{-\infty}^{\eta}{d\eta'}a^4(\eta')\bigg[G^>_k(\eta,\eta')-G^<_k(\eta,\eta')\bigg]\int\frac{d^3{\vec p}}{(2\pi)^3}G^>_p(\eta',\eta')\nonumber \\
&&\times\int_{-\infty}^{\eta'}{d\eta''}a^4(\eta'')\bigg[G^>_k(\eta',\eta'')G^>_k(\eta,\eta'')-G^<_k(\eta',\eta'')G^<_k(\eta,\eta'')\bigg]\ln{(-\kappa\eta'')}.
\label{1stsum}
\end{eqnarray}
The sum of \eqref{2-cross} and of the second part of \eqref{loop-cr} (the part with only the second term kept in the curly brackets) gives
\begin{eqnarray}
&&{6\lambda{\delta_m}H^2\over 4\pi^2}\int\frac{d^3{\vec k}}{(2\pi)^3}{e^{i\vec k\cdot(\vec x-\vec y)}}\int_{-\infty}^{\eta}{d\eta'}a^4(\eta')\bigg[G^>_k(\eta,\eta')-G^<_k(\eta,\eta')\bigg]\nonumber \\
&&\times\int_{-\infty}^{\eta'}{d\eta''}a^4(\eta'')\bigg[G^>_k(\eta',\eta'')G^>_k(\eta,\eta'')-G^<_k(\eta',\eta'')G^<_k(\eta,\eta'')\bigg]\ln{(-\kappa\eta'')}\;.
\label{2ndsum}
\end{eqnarray}
Finally, putting together \eqref{1stsum} and \eqref{2ndsum}, we get
\begin{eqnarray}
&&-{9\lambda^2H^4\over 8\pi^4}\int\frac{d^3{\vec k}}{(2\pi)^3}{e^{i\vec k\cdot(\vec x-\vec y)}}\int_{-\infty}^{\eta}{d\eta'}a^4(\eta')\bigg[G^>_k(\eta,\eta')-G^<_k(\eta,\eta')\bigg]\ln{(-\kappa\eta')}\nonumber \\
&&~~~\times\int_{-\infty}^{\eta'}{d\eta''}a^4(\eta'')\bigg[G^>_k(\eta',\eta'')G^>_k(\eta,\eta'')-G^<_k(\eta',\eta'')G^<_k(\eta,\eta'')\bigg]\ln{(-\kappa\eta'')}.
\label{fig.2}
\end{eqnarray}
To obtain \eqref{1stsum}, \eqref{2ndsum} and \eqref{fig.2}, we used the result \eqref{1loopreg} for the loop integral and \eqref{mc} for the mass counterterm (with $\mu_{UV}=H$). Similar to the previous section, in order to perform the integrals over the time variables, we need to change the integration variables: from $\eta''$ to $z''\equiv k\eta''$ and from $\eta'$ to $z'\equiv k\eta'$. Consequently, we also need to rewrite the time-dependent logarithms 
\begin{eqnarray}
\ln{(-\kappa\eta')}\ln{(-\kappa\eta'')}d\eta''d\eta'=\bigg(\ln{(-z')}-\ln{{k\over\kappa}}\bigg)\bigg(\ln{(-z'')}-\ln{{k\over\kappa}}\bigg){dz''dz'\over k^2}.
\label{split2}
\end{eqnarray}
Just as in \eqref{1looplate}, there aren't any contributions coming from the lower limits of the time integrals in \eqref{fig.2}. To evaluate the upper limit contributions we expand the Wightman functions ($\eta\to0$, $\eta'\to0$ and $\eta''\to0$) 
\begin{eqnarray}
&&a^4(\eta')a^4(\eta'')\bigg[G^>_k(0,\eta')-G^<_k(0,\eta')\bigg]\bigg[G^>_k(\eta',\eta'')G^>_k(0,\eta'')-G^<_k(\eta',\eta'')G^<_k(0,\eta'')\bigg]\nonumber\\&&~~~=-{1\over 9H^2 k^3\eta'\eta''}\bigg(1+\mathcal{O}(k^2\eta'^2, k^2\eta''^2)\bigg)=-{1\over 9H^2 k z' z''}\bigg(1+\mathcal{O}(z'^2, z''^2)\bigg).
\label{expand2}
\end{eqnarray}
Using \eqref{split2} and \eqref{expand2} to perform the time integrals, we find that at late times the long-wavelength part of \eqref{fig.2}, with coinciding external spatial points ($\vec x=\vec y$), goes as
\begin{eqnarray}
{\lambda^2 H^2\over 32\pi^6}\int_{\kappa}^{-1/\eta}\frac{d{k}}{k}\bigg(\ln^2{{k\over\kappa}}-\ln{{k\over\kappa}}\ln{(-k\eta)}+{1\over 4}\ln^2{(-k\eta)}\bigg)\ln^2{(-k\eta)}=-{\lambda^2 H^2\over 240\pi^6}\ln^5{(-\kappa\eta)}={\lambda^2 H^7\over 240\pi^6}(t-t_0)^5.
\label{one}
\end{eqnarray}
\subsubsection{Snowman diagram}
Next, we consider the diagrams on Fig.3: the snowman diagram and the corresponding counterterm diagram with the mass counterterm insertion in its loop.

\setlength{\unitlength}{1mm}
\thicklines
\begin{picture}(250,40)
\put(0,15){\circle*{2}}
\put(-4,11){$(\vec{x},t)$}
\put(0,15){\line(1,0){15}}
\put(15,15){\circle*{2}}
\put(14,11){$t'$}
\put(15,20){\circle{10}}
\put(15,25){\circle*{2}}
\put(14,21){$t''$}
\put(15,30){\circle{10}}
\put(15,15){\line(1,0){15}}
\put(30,15){\circle*{2}}
\put(26,11){$(\vec{y},t)$}
\put(30,5){Fig. 3}
\put(40,15){\circle*{2}}
\put(40,11){}
\put(40,15){\line(1,0){15}}
\put(55,15){\circle*{2}}
\put(55,11){}
\put(55,20){\circle{10}}
\put(53,28){$\delta_m$}
\put(53,24){$\bigotimes$}
\put(53,20){}
\put(54,15){\line(1,0){15}}
\put(69,15){\circle*{2}}
\put(69,11){}
\end{picture}
The sum of these diagrams gives
\begin{eqnarray*}
&&(-i)^2{\lambda\over 4}\int\frac{d^3{\vec k}}{(2\pi)^3}{e^{i\vec k\cdot(\vec x-\vec y)}}\int_{-\infty}^{\eta}{d\eta'}a^4(\eta')\bigg[(G^>_k(\eta,\eta'))^2-(G^<_k(\eta,\eta'))^2\bigg]\nonumber\\
&&~~~\times\int_{-\infty}^{\eta'}{d\eta''}a^4(\eta'')\int\frac{d^3{\vec p}}{(2\pi)^3}\bigg[(G^>_p(\eta',\eta''))^2-(G^<_p(\eta',\eta''))^2\bigg]\bigg\{{288\over 2}\cdot{\lambda\over 4}\int\frac{d^3{\vec \ell}}{(2\pi)^3}G^>_{\ell}(\eta'',\eta'')+24\cdot{{\delta_m}\over 2}\bigg\}.
\nonumber\\
\label{snowman}
\end{eqnarray*} 
Using \eqref{1loopreg} and the result \eqref{mc} for the mass counterterm (with $\mu_{UV}=H$), the above expression becomes  
\begin{eqnarray}
&&{9\lambda^2H^2\over 4\pi^2}\int\frac{d^3{\vec k}}{(2\pi)^3}{e^{i\vec k\cdot(\vec x-\vec y)}}\int_{-\infty}^{\eta}{d\eta'}a^4(\eta')\bigg[(G^>_k(\eta,\eta'))^2-(G^<_k(\eta,\eta'))^2\bigg]\nonumber\\
&&~~~\times\int_{-\infty}^{\eta'}{d\eta''}a^4(\eta'')\int\frac{d^3{\vec p}}{(2\pi)^3}\bigg[(G^>_p(\eta',\eta''))^2-(G^<_p(\eta',\eta''))^2\bigg]\ln{(-\kappa\eta'')}.
\label{snowman1}
\end{eqnarray}
\end{widetext}
Unlike in \eqref{1loop}, where the loop momentum $\vec\ell$ is associated with only one time variable, $\eta'$, in \eqref{snowman1} the momentum $\vec p$ appears in the products with $\eta'$ as well as $\eta''$. This means that if we want to regulate the integral over $\vec p$ with the UV-cutoff set at a fixed physical momentum scale $\Lambda_{UV}$, there is ambiguity in choosing the comoving cutoff scale: should it be $a(\eta')\Lambda_{UV}$ or $a(\eta'')\Lambda_{UV}$? One way to overcome this problem is to perform the $\eta''$ integration before the $\vec p$ integration. 

So we start by evaluating the integral over $\eta''$. This time variable enters the exponents in \eqref{snowman1} as $(\pm p\eta'')$; hence, we change the integration variable from $\eta''$ to $z''\equiv p\eta''$ and express the time-dependent logarithm in terms of $z''$: 
\begin{eqnarray}
\ln{(-\kappa\eta'')}d\eta''=\bigg(\ln{(-z'')}-\ln{{p\over\kappa}}\bigg){dz''\over p}.
\end{eqnarray}
Using the late-time expansion of the Wightman functions ($\eta\to0$, $\eta'\to0$ and $\eta''\to0$)
\begin{widetext}
\begin{eqnarray*}
&&a^4(\eta')a^4(\eta'')\bigg[(G^>_k(0,\eta'))^2-(G^<_k(0,\eta'))^2\bigg]\bigg[(G^>_p(\eta',\eta''))^2-(G^<_p(\eta',\eta''))^2\bigg]\nonumber\\&&~~~=-{1\over 9 \eta'\eta''k^3p^3}\bigg(1+\mathcal{O}(k^2\eta'^2, p^2\eta'^2, p^2\eta''^2)\bigg)=-{1\over 9 \eta'z''k^3p^2}\bigg(1+\mathcal{O}(k^2\eta'^2, p^2\eta'^2, z''^2)\bigg),
\end{eqnarray*}
\end{widetext}
and integrating over $z''$, we find the leading late-time behavior of the integrand of the integral over $\eta'$ in \eqref{snowman1}:
\begin{eqnarray}
{H^2\over 9k^3\eta'}\int{d^3{\vec p}\over(2\pi)^3}{1\over p^3}\left[{1\over2}\ln^2{(-p\eta')}-\ln{(-\kappa\eta')}\ln{(-p\eta')}\right].\nonumber\\
\label{smmom}
\end{eqnarray}
It is convenient to split this momentum integral (UV- and IR-regulated) in the following way,
\begin{eqnarray}
\!\!\!\!\!\!\!\!\int_{\kappa}^{a(\eta')\Lambda_{UV}}{d^3{\vec p}\over(2\pi)^3}=\int_{\kappa}^{-1/\eta'}{d^3{\vec p}\over(2\pi)^3}+\int_{-1/\eta'}^{\Lambda_{UV}\over -H\eta'}{d^3{\vec p}\over(2\pi)^3}.\nonumber\\
\label{split}
\end{eqnarray}
At late times the first term on the right side dominates, so the leading late-time behavior of \eqref{smmom} is 
\begin{eqnarray*}
&&{H^2\over 9k^3\eta'}\int_{\kappa}^{-1/\eta'}{d^3{\vec p}\over(2\pi)^3}{1\over p^3}\bigg[{1\over2}\ln^2{(-p\eta')}-\ln{(-\kappa\eta')}\ln{(-p\eta')}\bigg]\nonumber\\&&~~={H^2\over 54k^3\eta'\pi^2}\ln^3{(-\kappa\eta')}.
\end{eqnarray*}
To perform the $\eta'$ integral, we switch the integration variable from $\eta'$ to $z'\equiv k\eta'$, and hence split the above logarithm
\begin{eqnarray*}
\ln^3{(-\kappa\eta')}{d\eta'\over\eta'}=\left(\ln{(-z')}-\ln{{k\over\kappa}}\right)^3{dz'\over z'}.
\end{eqnarray*}
Evaluating the integral over $z'$ and, subsequently, over $\vec k$, we find the dominant late-time behavior of the long-wavelength part of \eqref{snowman1} (with $\vec x=\vec y$)   

\begin{widetext}
\begin{eqnarray}
&&{\lambda^2 H^2\over48\pi^6}\!\!\!\!\int_{\kappa}^{-1/\eta}\frac{d{k}}{k}\bigg({3\over2}\ln^2{\bigg({k\over\kappa}\bigg)}\ln{(-k\eta)}-\ln{\bigg({k\over\kappa}\bigg)}\ln^2{(-k\eta)}-\ln^3{\bigg({k\over\kappa}\bigg)}+{1\over4}\ln^3{(-k\eta)}\bigg)\ln{(-k\eta)}\nonumber\\
&&~~=-{\lambda^2 H^2\over240\pi^6}\ln^5{(-\kappa\eta)}={\lambda^2 H^7\over240\pi^6}(t-t_0)^5.
\label{two}
\end{eqnarray}

\subsubsection{Sunset diagram}
For the field contractions that correspond to the sunset diagram (Fig.4) we obtain the following expression,
\begin{eqnarray}
&&2(192){1\over 2}\bigg({-i\lambda\over 4}\bigg)^2\int\frac{d^3{\vec k}}{(2\pi)^3}{e^{i\vec k\cdot(\vec x-\vec y)}}\int_{-\infty}^{\eta}{d\eta'}a^4(\eta')\bigg[G^>_k(\eta,\eta')-G^<_k(\eta,\eta')\bigg]\int_{-\infty}^{\eta'}{d\eta''}a^4(\eta'')\nonumber\\&&~~~\times\int\frac{d^3{\vec q}}{(2\pi)^3}\frac{d^3{\vec p}}{(2\pi)^3}\frac{d^3{\vec \ell}}{(2\pi)^3}(2\pi)^3\delta^3(\vec k+\vec p+\vec l+\vec q)\nonumber\\&&~~~\times\bigg[G^>_k(\eta,\eta'')G^>_p(\eta',\eta'')G^>_{\ell}(\eta',\eta'')G^>_q(\eta',\eta'')-G^<_k(\eta,\eta'')G^<_p(\eta',\eta'')G^<_{\ell}(\eta',\eta'')G^<_q(\eta',\eta'')\bigg].
\label{sunset}
\end{eqnarray}
As in the case of the snowman diagram, the loop momenta in \eqref{sunset} is associated with two time variables, $\eta'$ and $\eta''$, which makes it unclear how to choose the comoving UV-cutoff. Hence, we will integrate over $\eta''$ before integrating over momenta. Expanding the Wightman functions in \eqref{sunset} ($\eta\to0$, $\eta'\to0$ and $\eta''\to0$) 
\begin{eqnarray}
&&a^4(\eta')a^4(\eta'')\bigg[G^>_k(0,\eta')-G^<_k(0,\eta')\bigg]\nonumber\\&&~~~\times\bigg[G^>_k(0,\eta'')G^>_p(\eta',\eta'')G^>_{\ell}(\eta',\eta'')G^>_q(\eta',\eta'')-G^<_k(0,\eta'')G^<_p(\eta',\eta'')G^<_{\ell}(\eta',\eta'')G^<_q(\eta',\eta'')\bigg]\nonumber\\&&=-{H^2\over72k^3\eta'\eta''}\bigg({k^3+p^3+\ell^3+q^3\over p^3\ell^3q^3}\bigg)+\cdots,
\end{eqnarray}
and changing the integration variable from $\eta''$ to $z''\equiv(k+p+\ell+q)\eta''$, since that's how it appears in the exponents in \eqref{sunset}, we obtain 
\begin{eqnarray}
&&-{H^2\over72k^3\eta'}\int\frac{d^3{\vec q}}{(2\pi)^3}\frac{d^3{\vec p}}{(2\pi)^3}\frac{d^3{\vec \ell}}{(2\pi)^3}(2\pi)^3\delta^3(\vec k+\vec p+\vec l+\vec q){k^3+p^3+\ell^3+q^3\over p^3\ell^3q^3}\ln{[-\eta'(k+p+\ell+q)]}\nonumber\\&&=-{H^2\over72k^3\eta'}\int\frac{d^3{\vec p}}{(2\pi)^3}\frac{d^3{\vec \ell}}{(2\pi)^3}{k^3+3q^3\over p^3\ell^3q^3}\ln{[-\eta'(k+p+\ell+q)]},
\label{sunsetmom}
\end{eqnarray}
where $q=\lvert\vec k+\vec p+\vec\ell\rvert$.
\\\
\end{widetext}

\setlength{\unitlength}{1mm}
\thicklines
\begin{picture}(90,30)
\put(3,30){\circle*{2}}
\put(0,25){$(\vec{x},t)$}
\put(3,30){\line(1,0){20}}
\put(23,30){\circle*{2}}
\put(20,25){$t'$}
\put(23,30){\line(1,0){14}}
\put(37,30){\circle*{2}}
\put(37,25){$t''$}
\put(37,30){\line(1,0){20}}
\put(57,30){\circle*{2}}
\put(54,25){$(\vec{y},t)$}
%\put(30,30){\circle{60}}
%\qbezier(20,30)(30,40)(20,30)
\put(30,30){\circle{50}}
\put(25,15){Fig. 4}
\end{picture}
As in \eqref{split}, we split the momentum integrals in the following way 
\begin{eqnarray*}
&&\int_{\kappa}^{a(\eta')\Lambda_{UV}}\frac{d^3{\vec p}}{(2\pi)^3}\int_{\kappa}^{a(\eta')\Lambda_{UV}}\frac{d^3{\vec \ell}}{(2\pi)^3}\nonumber \\
&&=\int_{\kappa}^{-1/\eta'}\frac{d^3{\vec p}}{(2\pi)^3}\int_{\kappa}^{-1/\eta'}\frac{d^3{\vec \ell}}{(2\pi)^3}\nonumber \\
&&+\int_{-1/\eta'}^{\Lambda_{UV}\over -H\eta'}\frac{d^3{\vec p}}{(2\pi)^3}\int_{-1/\eta'}^{\Lambda_{UV}\over -H\eta'}\frac{d^3{\vec \ell}}{(2\pi)^3}\nonumber\\
&&+2\int_{\kappa}^{-1/\eta'}\frac{d^3{\vec p}}{(2\pi)^3}\int_{-1/\eta'}^{\Lambda_{UV}\over -H\eta'}\frac{d^3{\vec \ell}}{(2\pi)^3}.
\end{eqnarray*}
We can see that at the late-time limit the leading contribution comes from the first term on the right-hand side of the equality. Also, since our goal is to evaluate the long-wavelength part of \eqref{sunset} (i.e., $-k\eta$ is small), the $k^3$ term in the numerator of \eqref{sunsetmom} can be neglected. Hence, the leading behavior of \eqref{sunsetmom} is   
\begin{eqnarray}
&&-{H^2\over24k^3\eta'}\int_{\kappa}^{-1/\eta'}\frac{d^3{\vec p}}{(2\pi)^3}\int_{\kappa}^{-1/\eta'}\frac{d^3{\vec \ell}}{(2\pi)^3}{\ln{[-\eta'(p+\ell)]}\over p^3\ell^3}\nonumber\\
&&=-{H^2\over96\pi^4k^3\eta'}\int_{\alpha}^1\frac{dy_1}{y_1}\int_{\alpha}^1\frac{dy_2}{y_2}\,{\ln{(y_1+y_2)}},
\label{mom2}
\end{eqnarray}      
where $y_1\equiv-p\eta'$, $y_2\equiv-\ell\eta'$ and ${\alpha}\equiv-\kappa\eta'$. To perform these integrals, we split the logarithm in the following way,
\begin{eqnarray}
&&\int_{\alpha}^1\frac{dy_1}{y_1}\int_{\alpha}^1\frac{dy_2}{y_2}\,\ln{(y_1+y_2)}\nonumber \\
&&~=\int_{\alpha}^1\frac{dy_1}{y_1}\bigg[\int_{\alpha}^1\frac{dy_2}{y_2}\,\ln{y_1}+\int_{\alpha}^1\frac{dy_2}{y_2}\,\ln{\bigg(1+{y_2\over y_1}\bigg)}\bigg]\nonumber\\
&&~={1\over2}\ln^3{\alpha}+\int_{\alpha}^1\frac{dy_1}{y_1}\int_{0}^1\frac{dy_2}{y_2}\,\ln{\bigg(1+{y_2\over y_1}\bigg)}.
\label{mom3}
\end{eqnarray}
In the integral over $y_2$ in the second term we set $\alpha=0$ because there is no divergence at $y_2=0$: it is an integral representation of the dilogarithm function
\begin{eqnarray}
\text{Li}_2(x)=-\int_0^1{\ln{(1-xt)}\over t}{dt}\;,
\label{dilog}
\end{eqnarray}
so \eqref{mom3} can be written as
\begin{eqnarray}
&&\int_{\alpha}^1\frac{dy_1}{y_1}\int_{\alpha}^1\frac{dy_2}{y_2}\,\ln{(y_1+y_2)}\nonumber \\
&&={1\over2}\ln^3{\alpha}-\int_{\alpha}^1\text{Li}_2\bigg(\!\!-{1\over y_1}\bigg)\frac{dy_1}{y_1}\;.
\label{mom4}
\end{eqnarray}
In order to deal with the integral that involves the dilogarithm function, we will use its the large argument expansion \cite{Wood}: for $x\gg1$ (i.e., $y_1\ll1$),
\begin{eqnarray}
\text{Li}_2(-x)={\pi^2\over6}-{1\over2}\ln^2{x}+\mathcal{O}(1/x)\;. 
\label{largarg}
\end{eqnarray}
Replacing the dilogarithm by the logarithmic term of this expansion, we obtain the leading behavior of \eqref{mom4}
\begin{eqnarray}
\to{1\over2}\ln^3{\alpha}+{1\over2}\int_{\alpha}^1{\ln^2{y_1}\over y_1}dy_1={1\over3}\ln^3{(-\kappa\eta')}\;.
\end{eqnarray}
To perform the $\eta'$ integral, we switch the variable from $\eta'$ to $z'\equiv k\eta'$, and hence split the above logarithm
\begin{eqnarray}
\ln^3{(-\kappa\eta')}{d\eta'\over\eta'}=\bigg(\ln{(-z')}-\ln{{k\over\kappa}}\bigg)^3{dz'\over z'}\;.
\label{split3}
\end{eqnarray}
Using \eqref{mom3} and \eqref{split3} to integrate \eqref{mom2} over $\eta'$, and then integrating over $\vec k$, we obtain that at late times the long-wavelength part of \eqref{sunset}, with $\vec x=\vec y$, goes as      
\begin{widetext}
\begin{eqnarray}
{\lambda^2 H^2\over 16\pi^6}&&\int_{\kappa}^{-1/\eta}\frac{d{k}}{k}\bigg({1\over2}\ln^2{\bigg({k\over\kappa}\bigg)}\ln{(-k\eta)}-{1\over3}\ln{\bigg({k\over\kappa}\bigg)}\ln^2{(-k\eta)}-{1\over3}\ln^3{\bigg({k\over\kappa}\bigg)}+{1\over 12}\ln^3{(-k\eta)}\bigg)\ln{(-k\eta)}\nonumber\\&&=-{\lambda^2 H^2\over240\pi^6}\ln^5{(-\kappa\eta)}={\lambda^2 H^7\over240\pi^6}(t-t_0)^5\;.
\label{three}
\end{eqnarray}

We would like to say a few words about the counterterm diagrams shown on Fig.5.

\setlength{\unitlength}{1mm}
\thicklines
\begin{picture}(250,30)
\put(0,15){\circle*{2}}
\put(-4,11){$(\vec{x},t)$}
\put(0,15){\line(1,0){15}}
\put(15,14){$\bigotimes$}
\put(18,11){$t'$}
\put(18,15){\line(1,0){15}}
\put(33,15){\circle*{2}}
\put(30,11){$(\vec{y},t)$}
\put(17,20){\circle{10}}
\put(15,18.5){$\delta_{\lambda}$}
\put(40,15){\circle*{2}}
\put(40,11){}
\put(40,15){\line(1,0){15}}
\put(55,14.5){$\bigotimes$}
\put(55,11){}
\put(55,18.5){$\delta_1$}
\put(58,15){\line(1,0){15}}
\put(73,15){\circle*{2}}
\put(69,11){}
\put(80,15){\circle*{2}}
\put(80,11){}
\put(80,15){\line(1,0){15}}
\put(95,14){$\bigotimes$}
\put(95,11){}
\put(95,18.5){$\delta_2$}
\put(98,15){\line(1,0){15}}
\put(113,15){\circle*{2}}
\put(110,11){}
\put(120,15){\circle*{2}}
\put(120,11){}
\put(120,15){\line(1,0){15}}
\put(135,14.5){$\bigotimes$}
\put(140,11){}
\put(135,18.5){$\delta_m|_{\lambda^2}$}
\put(138,15){\line(1,0){15}}
\put(154,15){\circle*{2}}
\put(151,11){}
\put(75,5){Fig. 5}
\end{picture}
\end{widetext}
For $t\to\infty$, the Fourier transform of the first diagram is proportional to $\bigl(\ln^2{(-k\eta)}/k^3\bigr)\delta_\lambda$, so its contribution to $\langle\phi^2(\vec x,t)\rangle_L$ is proportional to $\lambda^2t^3$; the Fourier transform of the second diagram is proportional to $\bigl(\ln{(-k\eta)}/k^3\bigr)\delta_m|_{\lambda^2}$, and its contribution to $\langle\phi^2(\vec x,t)\rangle_L$ is proportionall to $\lambda^2t^2$. The Fourier transforms of the second and third diagrams don't have any late-time divergencies, because only two powers of the scale factor enter the vertex time integration, unlike all other graphs, which have four powers of the scale factor at each vertex. Straightforward calculation of these diagrams shows that one of them is proportional to $\delta_1(k^2\eta^2-1)/k^3$, and the other one --- to $\delta_2(k^2\eta^2+3)/k^3$, so their contributions to $\langle\phi^2(\vec x,t)\rangle_L$ are proportional to $\lambda^2t$. Hence, we can see that the late-time contributions of the diagrams on Fig.5  are subdominant.        

Finally, we conclude that the leading late-time behavior of $\langle\phi^2(\vec x,t)\rangle_{\lambda^2,\,L}$ is given by the sum of \eqref{one}, \eqref{two} and \eqref{three},
\begin{eqnarray}
\langle\phi^2(\vec x,t)\rangle_{\lambda^2, L}\approx{\lambda^2 H^7\over 80\pi^6}(t-t_0)^5
\label{2point2}
\end{eqnarray}

\subsection{$\langle\phi^4(\vec x,t)\rangle$ to first order in $\lambda$}

At zeroth order in $\lambda$ the equal-time four-point function is the sum of products of the free theory two-point functions 
\begin{eqnarray}
&&\langle\phi(\vec x_1,t)\phi(\vec x_2,t)\phi(\vec x_3,t)\phi(\vec x_4,t)\rangle_{\lambda^0}\nonumber \\
&&=\langle 0|\phi(\vec x_1,t)\phi(\vec x_2,t)\phi(\vec x_3,t)\phi(\vec x_4,t)|0\rangle\nonumber\\
&&=G^{>}(x_1,x_2)G^{>}(x_3,x_4)+G^{>}(x_1,x_3)G^{>}(x_2,x_4)\nonumber\\&&~~+G^{>}(x_1,x_4)G^{>}(x_2,x_3). 
\end{eqnarray}
For coinciding spatial points this gives
\begin{eqnarray*}
\langle\phi^4(\vec x,t)\rangle_{\lambda^0}=3\bigl(G^{>}(x,x)\bigr)^2=3\bigl(\langle\phi^2(\vec x,t)\rangle_{\lambda^0}\bigr)^2,
\end{eqnarray*}  
and hence, for $t\to\infty$, the long-wavelength part of $\langle\phi^4(\vec x,t)\rangle_{\lambda^0}$ can be obtained from \eqref{late0}:
\begin{eqnarray}
\langle\phi^4(\vec x,t)\rangle_{\lambda^0, L}={3H^6\over16\pi^4}(t-t_0)^2\;.
\label{4point0}
\end{eqnarray}
The first-order correction to the equal-time four-point function is given by
\begin{widetext}
\begin{eqnarray}  
&&\langle\phi(\vec x_1,t)\phi(\vec x_2,t)\phi(\vec x_3,t)\phi(\vec x_4,t)\rangle_{\lambda}\nonumber\\&&~~~=-i\int_{-\infty}^t dt'\;\langle0|T(\phi^+(\vec x_1,t)\phi^+(\vec x_2,t)\phi^+(\vec x_3,t)\phi^+(\vec x_4,t)[H_I^+(t')-H_I^-(t')])|0\rangle\nonumber\\&&~~~=\langle\phi(\vec x_1,t)\phi(\vec x_2,t)\rangle_{\lambda^0}\langle\phi(\vec x_3,t)\phi(\vec x_4,t)\rangle_{\lambda}+(\vec x_1\leftrightarrow\vec x_3,\,\vec x_2\leftrightarrow\vec x_4)\nonumber\\&&~~~~~~+\langle\phi(\vec x_1,t)\phi(\vec x_3,t)\rangle_{\lambda^0}\langle\phi(\vec x_2,t)\phi(\vec x_4,t)\rangle_{\lambda}+(\vec x_1\leftrightarrow\vec x_2,\,\vec x_3\leftrightarrow\vec x_4)\nonumber\\&&~~~~~~+\langle\phi(\vec x_1,t)\phi(\vec x_4,t)\rangle_{\lambda^0}\langle\phi(\vec x_2,t)\phi(\vec x_3,t)\rangle_{\lambda}+(\vec x_1\leftrightarrow\vec x_2,\,\vec x_4\leftrightarrow\vec x_3)\nonumber\\&&~~~~~~+\langle\phi(\vec x_1,t)\phi(\vec x_2,t)\phi(\vec x_3,t)\phi(\vec x_4,t)\rangle_{\lambda}^{\rm connected}.
\label{4point}
\end{eqnarray}
The first six terms are simply the products of the $\lambda$-correction to the two-point function, calculated in Section 3, and the free theory two-point function. The last term is the fully connected piece shown on Fig.6.

\setlength{\unitlength}{1mm}
\thicklines
\begin{picture}(45,45)
\put(5,40){\circle*{2}}
\put(-5.5,40){$(\vec{x_1},t)$}
\put(6,39){\line(1,-1){14}}
\put(21,24){\circle*{2}}
\put(20,19){$t'$}
\put(22,25){\line(1,1){14}}
\put(22,23){\line(1,-1){14}}
\put(20,23){\line(-1,-1){14}}
\put(37,40){\circle*{2}}
\put(38.5,40){$(\vec{x_2},t)$}
\put(5,8){\circle*{2}}
\put(-5.5,8){$(\vec{x_3},t)$}
\put(37,8){\circle*{2}}
\put(38.5,8){$(\vec{x_4},t)$}
\put(19,1){Fig. 6}
\end{picture}
\\\\\ Evaluating contractions that correspond to this piece, we obtain  
\begin{eqnarray}
&&24\bigg({-i\lambda\over4}\bigg)\int\frac{d^3{\vec k_1}}{(2\pi)^3}\frac{d^3{\vec k_2}}{(2\pi)^3}\frac{d^3{\vec k_3}}{(2\pi)^3}\frac{d^3{\vec k_4}}{(2\pi)^3}\,e^{i\vec k_1\cdot\vec x_1}e^{i\vec k_2\cdot\vec x_2}e^{i\vec k_3\cdot\vec x_3}e^{i\vec k_4\cdot\vec x_4}(2\pi)^3\delta^3(\vec k_1+\vec k_2+\vec k_3+\vec k_4)\nonumber\\&&~~~\times\int_{-\infty}^{\eta}{d\eta'}a^4(\eta')\bigg[G^>_{k_1}(\eta,\eta')G^>_{k_2}(\eta,\eta')G^>_{k_3}(\eta,\eta')G^>_{k_4}(\eta,\eta')-G^<_{k_1}(\eta,\eta')G^<_{k_2}(\eta,\eta')G^<_{k_3}(\eta,\eta')G^<_{k_4}(\eta,\eta')\bigg].\nonumber\\
\label{4pointc}
\end{eqnarray}
What is the late-time behavior of \eqref{4pointc}? As we mentioned before, when performing the integral over $\eta'$, there are no contributions coming from its lower limit because the time contour is deformed to have a small imaginary part: at the lower limit $\eta'=-\infty(1\pm i\epsilon)$, and the oscillatory exponents in the Wightman functions turn into damping exponents. To evaluate the contribution of the upper limit of the integral for the late-time case, we expand the integrand as follows
\begin{eqnarray}
&&a^4(\eta')\bigg[G^>_{k_1}(0,\eta')G^>_{k_2}(0,\eta')G^>_{k_3}(0,\eta')G^>_{k_4}(0,\eta')-G^<_{k_1}(0,\eta')G^<_{k_2}(0,\eta')G^<_{k_3}(0,\eta')G^<_{k_4}(0,\eta')\bigg]\nonumber\\&&={iH^4\over24\eta'}\bigg({1\over k_1^3k_2^3k_3^3}+{1\over k_1^3k_2^3k_4^3}+{1\over k_1^3k_3^3k_4^3}+{1\over k_2^3k_3^3k_4^3}\bigg)+\cdots,
\end{eqnarray}
where dots stand for the terms that go to zero as $\eta'\to0$, starting from the term that is linear in $\eta'$. Using this expansion we can obtain the leading late-time behavior of \eqref{4pointc}. Hence, in the case of coinciding spatial points, we have   
\begin{eqnarray}
&&{\lambda H^4\over4}\int\frac{d^3{\vec k_1}}{(2\pi)^3}\frac{d^3{\vec k_2}}{(2\pi)^3}\frac{d^3{\vec k_3}}{(2\pi)^3}\frac{d^3{\vec k_4}}{(2\pi)^3}e^{i(\vec k_1+\vec k_2+\vec k_3+\vec k_4)\cdot\vec x}(2\pi)^3\delta^3(\vec k_1+\vec k_2+\vec k_3+\vec k_4)\nonumber\\&&\times\bigg({1\over k_1^3k_2^3k_3^3}+{1\over k_1^3k_2^3k_4^3}+{1\over k_1^3k_3^3k_4^3}+{1\over k_2^3k_3^3k_4^3}\bigg)\ln{[-\eta(k_1+k_2+k_3+k_4)]}.
\end{eqnarray}
\end{widetext}
As previously explained, there is a reason why the argument of the logarithm, which we obtained from the integration of $1/\eta'$, is made dimensionless by the sum of the magnitudes of the momenta, and not by some other quantity (e.g., by $H$ or by any of the $k_n$ separately): the arguments of the exponents in the Wightman functions look like $\pm(k_1+k_2+k_3+k_4)(\eta-\eta')$, so in order to perform the time integration in \eqref{4pointc} we need to change the integration variable from $\eta'$ to $(k_1+k_2+k_3+k_4)\eta'$. Taking into account that all four terms in the parentheses produce identical momentum integrals and using the delta-function we get
\begin{widetext}
\begin{eqnarray}
&&{\lambda H^4}\int\frac{d^3{\vec k_1}}{(2\pi)^3}\frac{d^3{\vec k_2}}{(2\pi)^3}\frac{d^3{\vec k_3}}{(2\pi)^3}{\ln{[-\eta\lvert\vec k_1+\vec k_2+\vec k_3\rvert-\eta(k_1+k_2+k_3)]}\over k_1^3k_2^3k_3^3}
\label{coinc}
\end{eqnarray}
\end{widetext}
We would like to find the long-wavelength part of the \eqref{coinc}, i.e., the part coming from the modes with physical momenta much less than the Hubble scale, $-k_n\eta\ll1$. Since
\begin{eqnarray}
\vert \vec k_1+\vec k_2+\vec k_3\rvert\leq k_1+k_2+k_3,
\end{eqnarray}
in evaluating the leading part of \eqref{coinc} for small momenta, we can neglect $\vert\vec k_1+\vec k_2+\vec k_3\rvert$. Hence, we arrive at the following expression that we need to calculate,  
\begin{eqnarray}
\!\!\!\!{\lambda H^4\over8\pi^6}\int_{\alpha}^1\frac{dy_1}{y_1}\int_{\alpha}^1\frac{dy_2}{y_2}\int_{\alpha}^1\frac{dy_3}{y_3}\,{\ln{(y_1+y_2+y_3)}},
\label{triple}
\end{eqnarray}
where $y_n\equiv-k_n\eta$ and ${\alpha}\equiv-\kappa\eta$, with $\kappa$ being some IR cutoff for the comoving momenta $k_n$. In order to integrate over $y_3$, let us split the logarithm just as we did in the previous section:
\begin{eqnarray}
&&\int_{\alpha}^1\frac{dy_3}{y_3}\,\ln{(y_1+y_2+y_3)}\nonumber \\
&&=\int_{\alpha}^1\frac{dy_3}{y_3}\,\ln{(y_1+y_2)}+\int_{0}^1\frac{dy_3}{y_3}\,\ln{\bigg(1+{y_3\over y_1+y_2}\bigg)}
\nonumber\\
&&=-\ln{\alpha}\ln{(y_1+y_2)}-\text{Li}_2\bigg(\!\!-{1\over y_1+y_2}\bigg),
\label{y_3}
\end{eqnarray}
where on the third line we used the integral representation of the dilogarithm function \eqref{dilog}. Similarly, splitting the $\ln{(y_1+y_2)}$ to perform the integration over $y_2$, we obtain 
\begin{eqnarray}
&&-\int_{\alpha}^1\frac{dy_1}{y_1}\int_{\alpha}^1\frac{dy_2}{y_2}\bigg[\ln{\alpha}\ln{(y_1+y_2)}+\text{Li}_2\bigg(\!\!-{1\over y_1+y_2}\bigg)\bigg]\nonumber\\
&&=-\ln{\alpha}\int_{\alpha}^1\frac{dy_1}{y_1}\bigg[-\ln{\alpha}\ln{y_1}-\text{Li}_2\bigg(\!\!-{1\over y_1}\bigg)\bigg]\nonumber \\
&&~~-\int_{\alpha}^1\frac{dy_1}{y_1}\int_{\alpha}^1\text{Li}_2\bigg(\!\!-{1\over y_1+y_2}\bigg){\ dy_2\over y_2}\nonumber\\
&&=-\ln{\alpha}\bigg[{1\over2}\ln^3{\alpha}-\int_{\alpha}^1\text{Li}_2\bigg(\!\!-{1\over y_1}\bigg){dy_1\over y_1}\bigg]\nonumber \\
&&~~-\int_{\alpha}^1{dy_1\over y_1}\int_{\alpha}^1\text{Li}_2\bigg(\!\!-{1\over y_1+y_2}\bigg){\ dy_2\over y_2}\;.
\label{y2y1}
\end{eqnarray}
Using the large argument expansion \eqref{largarg} to replace the dilogarithms by the logarithmic term of this expansion, we obtain the leading behavior of \eqref{y2y1}
\begin{eqnarray}
-{1\over3}\ln^4{\alpha}+{1\over2}\int_{\alpha}^1{dy_1\over y_1}\int_{\alpha}^1\ln^2{(y_1+y_2)}{\ dy_2\over y_2}.
\label{part1}
\end{eqnarray}
The second term of this expression can be broken up in the following way,
\begin{eqnarray}
&&{1\over2y_1}\int_{\alpha}^1\ln^2{(y_1+y_2)}{\ dy_2\over y_2}\nonumber \\
&&=-{\ln{\alpha}\ln^2{y_1}\over2y_1}-{1\over y_1}\int_{\alpha}^1{\ln{y_2}\ln{(y_1+y_2)}\over y_1+y_2}dy_2\nonumber\\
&&=-{\ln{\alpha}\ln^2{y_1}\over2y_1}-{1\over3y_1}\ln^3{(1+y_1)}+{1\over3y_1}\ln^3{(y_1+\alpha)}\nonumber\\
&&~~-{1\over y_1}\int_{y_1+\alpha}^{y_1+1}{\ln{(1-{y_1\over y})}\ln{y}\over y}dy.
\label{y2}
\end{eqnarray}
The second term on the third line has no divergence at $y_1=0$, so its integral over $y_1$ gives an $\alpha$-independent constant. The integrals over $y_1$ of the first and the third term of this line give logarithm to the forth of the IR cutoff $\alpha$ (in the third term $\alpha$ can be set to zero):  
\begin{eqnarray}
{1\over12}\ln^4{\alpha}\label
{part2}
\end{eqnarray}
The term on the last line of \eqref{y2} can be written in terms of dilogarithm and trilogarithm functions:
\begin{eqnarray}
\!\!\!\!\!&&\int_{y_1+\alpha}^{y_1+1}{\ln{(1-{y_1\over y})}\ln{y}\over y}dy\nonumber \\
&&~=\int_{0}^{1/(y_1+1)}{\ln{(1-{y_1z})}\ln{z}\over z}dz\nonumber \\
&&~~~~-\int_{0}^{1/(y_1+\alpha)}{\ln{(1-{y_1z})}\ln{z}\over z}dz\nonumber\\
&&~=\int_{0}^{1}{\ln{(1-{y_1\over1+y_1}t)}\ln{t}\over t}dt\nonumber \\
&&~~~~-\ln{(1+y_1)}\!\!\int_{0}^{1}{\ln{(1-{y_1\over1+y_1}t)}\ln{t}\over t}dt\nonumber\\
&&~~~~-\int_{0}^{1}{\ln{(1-{y_1\over{\alpha}+y_1}t)}\ln{t}\over t}dt\nonumber \\
&&~~~~+\ln{(y_1+\alpha)}\!\!\int_{0}^{1}{\ln{(1-{y_1\over{\alpha}+y_1}t)}\ln{t}\over t}dt\nonumber\\
&&~=\text{Li}_3\bigg({y_1\over1+y_1}\bigg)+\ln{(1+y_1)}\text{Li}_2\bigg({y_1\over1+y_1}\bigg)\nonumber \\
&&~~~~-\text{Li}_3(1)-\ln{y_1}\text{Li}_2(1),
\label{dy}
\end{eqnarray}
where we used an integral representation of the trilogarithm function 
\begin{eqnarray}
\text{Li}_3(x)=\int_0^1{\ln{(1-xt)}\ln{t}\over t}{dt},
\end{eqnarray}
and also set $\alpha=0$ when obtaining the last equality. Expanding the dilogarithm and trilogarithm (divided by $y_1$) for small $y_1$
\begin{eqnarray}
&&{(1/y_1)}\text{Li}_3\bigg({y_1\over1+y_1}\bigg)=1+\mathcal{O}(y_1),\nonumber \\
&&({\ln{(1+y_1)}/ y_1})\text{Li}_2\bigg({y_1\over1+y_1}\bigg)=y_1+\mathcal{O}(y_1^2), 
\end{eqnarray}
we see that there is no divergence at $y_1=0$, so integrating them over $y_1$ results in an $\alpha$-independent constants. Integration of the terms (divided by $y_1$) on the last line of \eqref{dy} gives linear logarithm and logarithm squared of the IR cutoff $\alpha$. 

Finally, we can conclude that the leading part of the \eqref{triple} is the sum of the first term in \eqref{part1} and \eqref{part2}; hence,
\begin{eqnarray}
&&\langle\phi^4(\vec x,t)\rangle_{\lambda, L}^{\rm connected}
\approx-{\lambda H^4\over32\pi^6}\ln^4{\alpha}=-{\lambda H^4\over32\pi^6}\ln^4{(-\kappa\eta)}\nonumber\\
&&~=-{\lambda H^8\over 32\pi^6}(t-t_0)^4.
\label{4pointcl}
\end{eqnarray}
Writing \eqref{4point} for coinciding spatial points, 
\begin{eqnarray}
&&\langle\phi^4(\vec x,t)\rangle_{\lambda}=6\langle\phi^2(\vec x,t)\rangle_{\lambda^0}\langle\phi^2(\vec x,t)\rangle_{\lambda}+\langle\phi^4(\vec x,t)\rangle_{\lambda}^{\rm connected}\; ,\nonumber \\
\label{4pointx}
\end{eqnarray}
and using \eqref{4pointcl}, \eqref{late0} and \eqref{2point1}, we can deduce the leading late-time behavior of the long-wavelength part of \eqref{4pointx}:   
\begin{eqnarray}
\langle\phi^4(\vec x,t)\rangle_{\lambda, L}=-{3\lambda H^8\over 32\pi^6}(t-t_0)^4\;.
\label{4point1}
\end{eqnarray}

\section*{Appendix II: Full and free infrared reduced scalar fields and retarded Green's functions}

In paper \cite{Onemli} a very convenient technique for calculation of the leading infrared contributions to the correlation functions at different spacetime points is elaborated. Here we shall use this technique to present very simple calculations of the corresponding coefficients in the  expectation values $\langle \phi^2(\vec x,t)\rangle$ and $\langle \phi^4(\vec x,t)\rangle$. We shall use, as in \cite{Onemli}, the infrared reduced scalar field, where only the modes with $H < k < Ha(t)$ are retained. Then there are two kinds of infrared reduced scalar fields: the free field $\phi_0(\vec x,t)$, which satisfies the Klein-Gordon equation in the absence of the self-interaction, and the full infrared reduced scalar field $\phi(\vec x,t)$. These two fields are connected by the equation 
 \begin{eqnarray}
 \phi(t,\vec{x})&=&\phi_0(t,\vec {x})\nonumber \\
 &&-\int_0^tdt'a^{3}(t')\int d^{3}x\,G(t,\vec{x};t'\vec{x}')\frac{V'(\phi)}{1+\delta Z}.\nonumber\\
 \label{iteration}
 \end{eqnarray}
Here the Green's function $G$ satisfies the retarded boundary conditions, $Z$ is the renormalization constant of the scalar field 
and the potential $V$ includes the mass and coupling constant counterterms. Simple considerations \cite{Onemli} show  that the counterterms do not give contributions to the  leading infrared terms into the correlators. The leading infrared part of the retarded Green's function has the form 
\begin{equation}
G(t,\vec{x};t'\vec{x'})=\frac{1}{3H}\theta(t-t')\delta^{3}(\vec{x}-\vec{x}')\left[\frac{1}{a^{3}(t')}-\frac{1}{a^{3}(t)}\right].
\label{Green30}
\end{equation}
This expression in Eq. (\ref{iteration}) is multiplied by the integration measure $a^{3}(t')$. The first term in the square brackets, $a^{3}(t')/a^{3}(t') = 1$, contributes over the whole range of the integration; the second term, which is proportional to 
$a^{3}(t')/a^{3}(t)$, contributes significantly only for $t' \sim t$, and hence, is negligible in the  approximation we consider. Thus, the formula (\ref{iteration}) is boiled down to 
\begin{eqnarray}
\phi(t,\vec{x}) &=&\phi_0(t,\vec{x})-\frac{1}{3H}\int_0^t dt' V'(\phi(t',\vec{x}) \nonumber \\
&=&\phi_0(t,\vec{x})-\frac{\lambda}{3H}\int_0^t dt' \phi^3(t',\vec{x}).
\label{iter}
\end{eqnarray}
Solving Eq. (\ref{iter}) by iterations we obtain the following expression for the full scalar field $\phi(\vec x,t)$ expressed in terms of the free scalar field $\phi_0(\vec x,t)$ up to the second order in the coupling constant $\lambda$. In what follows we omit the argument $\vec{x}$ since it is the same in all the terms in our equations. Thus,
\begin{eqnarray}
\phi(t)&=&\phi_0(t) - \frac{\lambda}{3H}\int_0^tdt'\phi_0^3(t')\nonumber \\
&&+\frac{\lambda^2}{3H^2}\int_0^tdt'\phi_0^2(t')\int_0^{t'}dt''\phi_0^3(t'').
\label{iter1}
\end{eqnarray}
 \begin{widetext}
  Using this expression we obtain 
 \begin{eqnarray}
 \langle\phi^2(t)\rangle&=&\langle\phi_0^2(t)\rangle-\frac{\lambda}{3H}\left[\langle \phi_0(t)\int_0^tdt'\phi_0^3(t')\rangle+  
  \langle \int_0^tdt'\phi_0^3(t')\phi_0(t)\rangle\right]\nonumber \\
 &&+\frac{\lambda^2}{3H^2}\left[\langle \phi_0(t)\int_0^tdt'\phi_0^2(t')\int_0^{t'}dt''\phi_0^3(t'')\rangle\right.\langle \int_0^tdt'\phi_0^2(t')\int_0^{t'}dt''\phi_0^3(t'')\phi_0(t)\rangle\nonumber \\
 &&\left.+ \frac13\langle \int_0^tdt'\phi_0^3(t')\int_0^tdt''\phi_0^3(t'')\rangle\right].
  \label{average}
 \end{eqnarray}
 We shall also need the following formula
 \end{widetext}
 \begin{equation}
\langle\phi_0(t,\vec{x})\phi_0(t',\vec{x})\rangle = \frac{H^2}{4\pi^2}\ln(a(t'))=\frac{H^3t'}{4\pi^2} ,
\label{free}
\end{equation}
where   
\begin{equation}
t' \leq t,
\label{order}
\end{equation}   
Remarkably, in  the  formula (\ref{free}) the left-hand side contains both time moments $t$ and $t'$, while the right-hand side 
of (\ref{free}) depends only on the earlier moment $t'$. Note that in the formula (\ref{Green30}) for the Green's function both time moments $t$ and $t'$ are present, but when we integrate this Green's function with the corresponding measure, only the term depending on $a(t')$ gives the essential contribution. 

We are now in a position to make the necessary calculations. To calculate the $\lambda$-order contribution to the correlator we shall use Wick's theorem:
 \begin{widetext}
\begin{eqnarray}
 &&\langle\phi^2(t)\rangle_{\lambda}=-\frac{\lambda}{3H}\left[\langle \phi_0(t)\int_0^tdt'\phi_0^3(t')\rangle+  
  \langle \int_0^tdt'\phi_0^3(t')\phi_0(t)\rangle\right]\nonumber \\
 &&=-\frac{\lambda}{3H}\left[3\int_0^tdt'\langle\phi_0(t)\phi_0(t')\rangle\langle\phi_0^2(t')\rangle
 +3\int_0^tdt'\langle\phi_0^2(t')\rangle\langle\phi_0(t')\phi_0(t)\rangle\right].
 \label{Wick}
 \end{eqnarray}
  \end{widetext}
 Now using the formula (\ref{free}) and taking into account the order of the time moments (\ref{order}), we can rewrite the expression (\ref{Wick}) as follows:
\begin{eqnarray*}
 &&\langle\phi^2(t)\rangle_{\lambda}=-\frac{2\lambda}{H}\left(\frac{H^2}{4\pi^2}\right)^2\int_0^tdt'H^2t'^2=
 -\frac{\lambda H^5}{24\pi^4}t^3,
\label{Wick1}
\end{eqnarray*} 
 which coincides with the known result.
 Analogously, the $\lambda^2$-contribution to the correlator has the following form:
 \begin{widetext} 
 \begin{eqnarray}
 \langle\phi^2(t)\rangle_{\lambda^2}&=&\frac{2\lambda^2}{3H^2}\int_0^tdt'\int_0^{t'}dt''\bigg[6\langle\phi_0(t)\phi_0(t')\rangle\langle\phi_0(t')\phi_0(t'')\rangle
 \langle\phi_0^2(t'')\rangle\nonumber\\&&+3\langle\phi_0(t)\phi_0(t'')\rangle\langle\phi_0^2(t')\rangle\langle\phi_0^2(t'')\rangle+6\langle\phi_0(t)\phi_0(t'')\rangle(\langle\phi_0(t')\phi_0(t'')\rangle)^2\bigg]\nonumber \\
 &&+\frac{\lambda^2}{9H^2}\left[\int_0^tdt'\int_0^{t'}dt''\Big(6\left(\langle\phi_0(t')\phi_0(t'')\rangle\right)^3
 +9\langle\phi_0(t')\phi_0(t'')\rangle\langle\phi_0^2(t')\rangle\langle\phi_0^2(t'')\rangle\Big)\right.\nonumber \\
 &&\left.+\int_0^tdt'\int_{t'}^tdt''\Big(6(\langle\phi_0(t')\phi_0(t'')\rangle)^3
 +9\langle\phi_0(t')\phi_0(t'')\rangle\langle\phi_0^2(t')\rangle\langle\phi_0^2(t'')\rangle\Big)\right].
 \label{Wick2}
 \end{eqnarray}
Again, using the formula (\ref{free}) and taking into account the order of the time moments (\ref{order}), we can reduce  the expression (\ref{Wick2}) to simple integrals:
 \begin{eqnarray}
 \langle\phi^2(t)\rangle_{\lambda^2}&=&\frac{2\lambda^2}{3H^2}\left(\frac{H^2}{4\pi^2}\right)^3H^3\int_0^tdt'\int_0^{t'}dt''(6t't''^2+3t't''^2+6t''^3)\nonumber \\
 &&+\frac{\lambda^2}{9H^2}\left(\frac{H^2}{4\pi^2}\right)^3H^3\left[\int_0^tdt'\int_0^{t'}dt''(6t''^3+9t't''^2)+\int_0^tdt'\int_{t'}^{t}dt''(6t'^3+9t'^2t'')\right]= \frac{\lambda^2 H^7}{80\pi^6}t^5\;.
 \label{Wick3}
 \end{eqnarray} 
By this method we can also easily calculate the four-point correlator. To first order in $\lambda$,
\begin{eqnarray}
\langle\phi^4(t)\rangle&=&\langle\left(\phi_0(t)-\frac{\lambda}{3H}\int_0^tdt'\phi_0^3(t')\right)^4\rangle= 3\langle\phi_0^4(t)\rangle-\frac{4\lambda}{3H}\langle\phi_0^3(t)\int_0^tdt'\phi_0^3(t')
\rangle\nonumber \\
 &=&3(\langle\phi_0^2(t)\rangle)^2- \frac{4\lambda}{3H}\int_0^tdt'\Big(9\langle\phi_0(t)\phi_0(t')\rangle 
 \langle\phi_0^2(t)\rangle\langle\phi_0^2(t')\rangle
 +6(\langle\phi_0(t)\phi_0(t')\rangle)^3\Big)
 \nonumber \\
 &=&\frac{3H^6t^2}{16\pi^4}-\frac{4\lambda}{3H}\left(\frac{H^3}{4\pi^2}\right)^3\int_0^tdt'(9tt'^2+6t'^3)=\frac{3H^6t^2}{16\pi^4} - \frac{3\lambda H^8t^4}{32\pi^6},
\label{four}
\end{eqnarray} 
and for the $\lambda^2$-contribution we have 
 \begin{eqnarray}
 \langle \phi^4(t)\rangle_{\lambda^2}&=&\frac{4\lambda^2}{3H^2}\int_0^tdt'\int_0^{t'}dt''\langle\phi_0^3(t)\phi_0^2(t')\phi_0^3(t'')\rangle+
 \frac{2\lambda^2}{3H^2}\int_0^tdt'\int_0^{t}dt''\langle\phi_0^2(t)\phi_0^3(t')\phi_0^3(t'')\rangle\nonumber \\
 &=&\frac{4\lambda^2}{3H^2}\left(\frac{H^3}{4\pi^2}\right)^4\int_0^tdt'\int_0^{t'}dt''\bigg[18\langle\phi_0^2(t)\rangle
 \langle\phi_0(t)\phi_0(t')\rangle\langle\phi_0(t')\phi_0(t'')\rangle\langle\phi_0^2(t'')\rangle\nonumber \\
 &&+9\langle\phi_0^2(t)\rangle\langle\phi_0(t)\phi_0(t'')\rangle\langle\phi_0^2(t')\rangle\langle\phi_0^2(t'')\rangle+18\langle\phi_0^2(t)\rangle\langle\phi_0(t)\phi_0(t'')\rangle\langle\phi_0(t')\phi_0(t'')\rangle^2\nonumber \\
 &&+18\langle\phi_0(t)\phi_0(t')\rangle^2\langle\phi_0(t)\phi_0(t'')\rangle\langle\phi_0^2(t'')\rangle\nonumber \\
&&+36\langle\phi_0(t)\phi_0(t')\rangle\langle\phi_0(t)\phi_0(t'')\rangle^2\langle\phi_0(t')\phi_0(t'')\rangle+6\langle\phi_0(t)\phi_0(t'')\rangle^3\langle\phi_0^2(t')\rangle\bigg]\nonumber \\
&&+\frac{2\lambda^2}{3H^2}\left(\frac{H^3}{4\pi^2}\right)^4\int_0^tdt'\int_0^tdt''\bigg[9\langle\phi_0^2(t)\rangle\langle\phi_0^2(t')\rangle\langle\phi_0(t')\phi_0(t'')\rangle\langle\phi_0^2(t'')\rangle\nonumber\\
&&+6\langle\phi_0^2(t)\rangle\langle\phi_0(t')\phi_0(t'')\rangle^3+18\langle\phi_0(t)\phi_0(t')\rangle^2\langle\phi_0(t')\phi_0(t'')\rangle\langle\phi_0^2(t'')\rangle\nonumber \\
&&+18\langle\phi_0(t)\phi_0(t')\rangle\langle\phi_0(t)\phi_0(t'')\rangle\rangle\langle\phi_0^2(t')\rangle\rangle\langle\phi_0^2(t'')\rangle\nonumber \\
&&+18\langle\phi_0(t)\phi_0(t'')\rangle^2\langle\phi_0(t')\phi_0(t'')\rangle\langle\phi_0^2(t')\rangle+36\langle\phi_0(t)\phi_0(t')\rangle\langle\phi_0(t)\phi_0(t'')\rangle\langle\phi_0(t')\phi_0(t'')\rangle^2\bigg]\nonumber \\
 &&=\frac{53\lambda^2H^{10}}{960\pi^8}t^6.
 \label{four2}
 \end{eqnarray} 
 \end{widetext}
Using this method one can easily calculate the leading infrared contributions into various correlators (in coinciding points) up to higher orders in the coupling constant $\lambda$. The calculations are also easy for the coinciding spatial points but with different time coordinates.


\begin{thebibliography}{99}
\bibitem{Shirkov1}
N. N. Bogolyubov and D.V. Shirkov, {\it Renormgroup? It is very simple}, Priroda No 8, 3 (1984) (in Russian). 
\bibitem{Shirkov}
D.~V.~Shirkov,
 %``The Bogolyubov renormalization group in theoretical and mathematical physics,''
[arXiv:hep-th/9903073 [hep-th]].
\bibitem{Wilson}
K.~G.~Wilson and J.~B.~Kogut,
  %``The Renormalization group and the epsilon expansion,''
  Phys.\ Rept.\  {\bf 12}, 75 (1974).
\bibitem{Shirkov2}
D.~V.~Shirkov and V.~F.~Kovalev,
  %``Bogolyubov renormalization group and symmetry of solution in mathematical physics,''
  Phys.\ Rept.\  {\bf 352}, 219 (2001) [arXiv:hep-th/0001210 [hep-th]].
\bibitem{dynamical}
L.~Y.~Chen, N.~Goldenfeld and Y.~Oono,
  %``The Renormalization group and singular perturbations: Multiple scales, boundary layers and reductive perturbation theory,''
  Phys.\ Rev.\ E {\bf 54}, 376 (1996) [arXiv:hep-th/9506161 [hep-th]].
\bibitem{Emil}
E.~T.~Akhmedov, U.~Moschella and F.~K.~Popov,
  %``Characters of different secular effects in various patches of de Sitter space,''
  Phys.\ Rev.\ D {\bf 99}, no. 8, 086009 (2019) [arXiv:1901.07293 [hep-th]].
\bibitem{Prokopec:2007ak}
T.~Prokopec, N.~Tsamis and R.~Woodard,
%``Stochastic Inflationary Scalar Electrodynamics,''
Annals Phys. \textbf{323}, 1324-1360 (2008) [arXiv:0707.0847 [gr-qc]].  
\bibitem{vacuum}
N.~A.~Chernikov and E.~A.~Tagirov,
  %``Quantum theory of scalar fields in de Sitter space-time,''
  Ann.\ Inst.\ H.\ Poincare Phys.\ Theor.\ A {\bf 9}, 109 (1968).
\bibitem{vacuum1}
C.~Schomblond and P.~Spindel,
  %``Unicity Conditions of the Scalar Field Propagator Delta(1) (x,y) in de Sitter Universe,''
  Ann.\ Inst.\ H.\ Poincare Phys.\ Theor.\  {\bf 25}, 67 (1976).
\bibitem{vacuum2}
T.~S.~Bunch and P.~C.~W.~Davies,
  %``Quantum Field Theory in de Sitter Space: Renormalization by Point Splitting,''
  Proc.\ Roy.\ Soc.\ Lond.\ A {\bf 360}, 117 (1978).
\bibitem{grow}
A.~A.~Starobinsky,
  %``Dynamics of Phase Transition in the New Inflationary Universe Scenario and Generation of Perturbations,''
  Phys.\ Lett.\  {\bf 117B}, 175 (1982).
\bibitem{grow1}
A.~D.~Linde,
  %``Scalar Field Fluctuations in Expanding Universe and the New Inflationary Universe Scenario,''
  Phys.\ Lett.\  {\bf 116B}, 335 (1982).
\bibitem{grow2}
A.~Vilenkin and L.~H.~Ford,
  %``Gravitational Effects upon Cosmological Phase Transitions,''
  Phys.\ Rev.\ D {\bf 26}, 1231 (1982).
\bibitem{grow3}
L.~H.~Ford and A.~Vilenkin,
  %``Global Symmetry Breaking in Two-dimensional Flat Space-time and in De Sitter Space-time,''
  Phys.\ Rev.\ D {\bf 33}, 2833 (1986).
  \bibitem{grow4}
  B.~Allen and A.~Folacci,
  %``The Massless Minimally Coupled Scalar Field in De Sitter Space,''
  Phys.\ Rev.\ D {\bf 35}, 3771 (1987).
\bibitem{Star}
A.~A.~Starobinsky,
  %``Stochastic De Sitter (inflationary) Stage In The Early Universe,''
  Lect.\ Notes Phys.\  {\bf 246}, 107 (1986).
  \bibitem{Star-Yoko}
A.~A.~Starobinsky and J.~Yokoyama,
  %``Equilibrium state of a selfinteracting scalar field in the De Sitter background,''
  Phys.\ Rev.\ D {\bf 50}, 6357 (1994) [arXiv:astro-ph/9407016 [astro-ph]].
\bibitem{Woodard}
N.~C.~Tsamis and R.~P.~Woodard,
  %``Stochastic quantum gravitational inflation,''
  Nucl.\ Phys.\ B {\bf 724}, 295 (2005) [arXiv:gr-qc/0505115 [gr-qc]].
\bibitem{Tereza}
H.~Collins, R.~Holman and T.~Vardanyan,
  %``The quantum Fokker-Planck equation of stochastic inflation,''
  JHEP {\bf 1711}, 065 (2017) [arXiv:1706.07805 [hep-th]].
\bibitem{Burgess:2014eoa}
C.~P.~Burgess, R.~Holman, G.~Tasinato, and M.~Williams, %``EFT Beyond the Horizon: Stochastic Inflation and How Primordial Quantum Fluctuations Go Classical,'' 
JHEP {\bf 1503}, 090 (2015) [arXiv:1408.5002 [hep-th]].
\bibitem{Burgess:2015ajz}
C.~P.~Burgess, R.~Holman, and G.~Tasinato,
%``Open EFTs, IR effects \& late-time resummations: systematic corrections in stochastic inflation,''
 JHEP {\bf 1601}, 153 (2016) [arXiv:1512.00169 [gr-qc]].
\bibitem{Burgess}
C.~P.~Burgess, L.~Leblond, R.~Holman and S.~Shandera,
  %``Super-Hubble de Sitter Fluctuations and the Dynamical RG,''
  JCAP {\bf 1003}, 033 (2010) [arXiv:0912.1608 [hep-th]].
\bibitem{Schwinger}
J.~S.~Schwinger,
  %``Brownian motion of a quantum oscillator,''
  J.\ Math.\ Phys.\  {\bf 2}, 407 (1961).
\bibitem{Keldysh}
L.~V.~Keldysh,
  %``Diagram technique for nonequilibrium processes,''
  Zh.\ Eksp.\ Teor.\ Fiz.\  {\bf 47}, 1515 (1964)
  [Sov.\ Phys.\ JETP {\bf 20}, 1018 (1965)].
\bibitem{Bakshi:1962dv}
P.~M.~Bakshi and K.~T.~Mahanthappa, J.\ Math.\ Phys.\  {\bf 4} (1963) 1; P.~M.~Bakshi and K.~T.~Mahanthappa, J.\ Math.\ Phys.\  {\bf 4} (1963) 12.  
\bibitem{Jordan}
R.~D.~Jordan,
  %``Effective Field Equations for Expectation Values,''
  Phys.\ Rev.\ D {\bf 33}, 444 (1986).

\bibitem{Weinberg:2005vy}
S.~Weinberg,
%``Quantum contributions to cosmological correlations,''
Phys. Rev. D \textbf{72}, 043514 (2005) [arXiv:hep-th/0506236 [hep-th]].
\bibitem{On-Wood}
V.~K.~Onemli and R.~P.~Woodard,
  %``Superacceleration from massless, minimally coupled phi**4,''
  Class.\ Quant.\ Grav.\  {\bf 19}, 4607 (2002) [arXiv:gr-qc/0204065 [gr-qc]].
\bibitem{Brunier}
T.~Brunier, V.~K.~Onemli and R.~P.~Woodard,
  %``Two loop scalar self-mass during inflation,''
  Class.\ Quant.\ Grav.\  {\bf 22}, 59 (2005) [arXiv:gr-qc/0408080 [gr-qc]].
\bibitem{Kahiya}
E.~O.~Kahya and V.~K.~Onemli,
  %``Quantum Stability of a w < -1 Phase of Cosmic Acceleration,''
  Phys.\ Rev.\ D {\bf 76}, 043512 (2007) [arXiv:gr-qc/0612026 [gr-qc]].
\bibitem{Onemli}
V.~K.~Onemli,
  %``Vacuum Fluctuations of a Scalar Field during Inflation: Quantum versus Stochastic Analysis,''
  Phys.\ Rev.\ D {\bf 91}, 103537 (2015) [arXiv:1501.05852 [gr-qc]].
\bibitem{Collins:2011mz}
H.~Collins,
%``Primordial non-Gaussianities from inflation,''
[arXiv:1101.1308 [astro-ph.CO]].
\bibitem{Maldacena:2002vr}
J.~M.~Maldacena,
%``Non-Gaussian features of primordial fluctuations in single field inflationary models,''
JHEP \textbf{05}, 013 (2003) [arXiv:astro-ph/0210603 [astro-ph]].
\bibitem{Peskin:1995ev} 
M.~E.~Peskin and D.~V.~Schroeder, {\it An Introduction to Quantum Field Theory} (Addison-Wesley, 1995)
\bibitem{Xue:2011hm}
W.~Xue, K.~Dasgupta and R.~Brandenberger,
%``Cosmological UV/IR Divergences and de-Sitter Spacetime,''
Phys. Rev. D \textbf{83}, 083520 (2011),
[arXiv:1103.0285 [hep-th]].  
\bibitem{Senatore:2009cf}
L.~Senatore and M.~Zaldarriaga,
%``On Loops in Inflation,''
JHEP \textbf{12}, 008 (2010),
[arXiv:0912.2734 [hep-th]].
\bibitem{Wood}
D.~Wood, {\it The Computation of Polylogarithms} (Technical Report 15-92*, University of Kent, Canterbury, UK, 1992), [http://www.cs.kent.ac.uk/pubs/1992/110].
\end{thebibliography}
\end{document}